# Single-photon detection enabled by negative differential conductivity in moiré superlattices

Krystian Nowakowski[1+], Hitesh Agarwal[1+], Sergey Slizovskiy[2], Robin Smeyers[3], Xueqiao Wang[4], Zhiren Zheng[4], Julien Barrier[1], David Barcons Ruiz[1], Geng Li[1], Riccardo Bertini[1], Matteo Ceccanti[1], Iacopo Torre[1,5], Bert Jorissen[3], Antoine Reserbat-Plantey[1,6], Kenji Watanabe[7], Takashi Taniguchi[8], Lucian Covaci[3], Milorad V. Milošević[3], Vladimir Fal'ko[2], Pablo Jarillo-Herrero[4], Roshan Krishna Kumar[1,9]*, Frank H. L. Koppens[1,10]*

[1] ICFO - Institut de Ciències Fotòniques, The Barcelona Institute of Science and Technology, Av. Carl Friedrich Gauss 3, 08860 Castelldefels (Barcelona), Spain
[2] Department of Physics and Astronomy, The University of Manchester, M13 9PL Manchester, United Kingdom
[3] Department of Physics and NANOlab Center of Excellence, University of Antwerp, Groenenborgerlaan 171, 2020 Antwerp, Belgium
[4] Department of Physics, Massachusetts Institute of Technology, Cambridge, MA, USA
[5] Department of Physics, Universitat Politècnica de Catalunya - BarcelonaTech (UPC), Campus Nord, Building B4-B5, C. Jordi Girona, 1-3, 08034 Barcelona, Spain
[6] Université Côte d'Azur, CNRS, CRHEA, rue Bernard Grégory, 06560 Valbonne, France
[7] Research Center for Functional Materials, National Institute for Materials Science, Tsukuba 305-0044, Japan
[8] International Center for Materials Nanoarchitectonics, National Institute for Materials Science, Tsukuba 305-0044, Japan
[9] Catalan Institute of Nanoscience and Nanotechnology (ICN2), CSIC and BIST, Barcelona, Spain
[10] ICREA - Institucio Catalana de Recerca i Estudis Avancats, 08010 Barcelona, Spain
* Corresponding author
+ These authors contributed equally

## Abstract

**Detecting individual light quanta is essential for quantum information, space exploration, advanced machine vision, and fundamental science. Here, we introduce a novel single photon detection mechanism using highly photosensitive non-equilibrium electron phases in moiré materials. Using tunable bands in bilayer graphene/hexagonal-boron nitride superlattices, we engineer negative differential conductance and a sensitive bistable state capable of detecting single photons. Operating in this regime, we demonstrate single-photon counting at mid-infrared (11.3 µm) and visible wavelengths (675 nm) and temperatures up to 25 K. This detector offers new prospects for broadband, high-temperature quantum technologies with CMOS compatibility and seamless integration into photonic integrated circuits (PICs). Our analysis suggests the mechanism underlying our device operation originates from negative differential velocity, and represents an important milestone in the field of high-bias transport in two-dimensional moiré quantum materials.**

Introduction

Single-photon detectors (SPDs) play a crucial role in modern optoelectronics[1,2,3] with the demands for long-wavelength and high-temperature capabilities growing rapidly. The state-of-the-art devices based on single photon avalanche diodes (SPADs), superconducting nanowire single-photon detectors (SNSPDs) and transition edge sensors (TESs)[1,4,5,] exhibit high performance, but are limited by their wavelength and temperature operating ranges. In SPADs, their semiconductor bandgaps impose sharp cut-offs requiring complex up-conversion techniques for longer wavelengths[6,7]. SNSPDs can extend their range to mid-IR wavelengths but only using low $T_c$ materials below 1 K[8,,9,10,11], because their intrinsic superconducting gap makes high temperature[12,13,14] and long-wavelength operation mutually exclusive[8,15,9,10,11]. Hence, disruptive progress requires devices operating under fundamentally distinct principles.

2D materials are well suited for sensitive photodetection because their intrinsically low heat capacities[,16] make them responsive to slight photon-induced temperature changes[17,18,19,20,21,22]. Moiré superlattices represent a new era in 2D materials engineering with their correlated electron phases and tunable infrared band gaps offering exciting prospects for photodetection [23,24,25]. Additionally, the large lattice constant imposes peculiar quantum properties under high electric fields, in which carriers can be accelerated to their band velocity limit, or even spontaneously created inducing an electron-hole plasma[26] (Fig. 1A). Because of the small superlattice Brillouin zone, carriers may even be decelerated by electric fields, attaining a negative velocity. Indeed, high-field phenomena in superlattices produce distinct nonlinear electrical responses that underpin some of the most ground-breaking optoelectronic devices to date[27], but remain little explored in moiré materials.

In this work, we leverage the high-field properties of moiré superlattices to develop a so far unexplored mechanism for single-photon detection with a larger operating temperatures and wavelength range. By engineering their non-equilibrium electron phases[26] using tunable band structure of bilayer graphene, we reveal a regime of in-plane negative differential conductance (NDC) which makes the system capable of detecting single light quanta over a broad wavelength range. Our analysis suggests that the NDC originates from particle acceleration into regions of the Brillouin zone with negative velocity (shaded red in Fig. 1A) effectively deaccelerating charge carriers[28,29]. The operation principle of our moiré single photon detectors (MSPDs) bares strong similarities with SNSPDs[1,2], leveraging a sharp critical current behavior with exceptional photosensitivity, while the distinct underlying physical mechanism enables higher operating temperatures (>20 K) and long-wavelength capabilities[4] (> 10 μm, < 109 meV), owed to the low-energy band gaps intrinsic to moiré bilayer graphene under a moderate displacement field.

**High-field transport in moiré superlattices**

We studied two moiré-based superlattice devices (DBLG$_A$ and DBLG$_B$) consisting of bilayer graphene aligned to hexagonal boron nitride (hBN) which showed similar behaviours including sensitivity to single-photons (for fabrication details and transport characterization see methods and Supplementary Section 1). Fig. 1B plots the resistivity measured as a function of the gate voltage ($V_G$) and filling factor ($v$), exhibiting features expected for bilayer graphene aligned to hBN, including a peaked response at neutrality point, and additional peaks for hole and electron doping indicating the band edges[30]. Fig. 1C plots the current-voltage (IV) characteristics of DBLG$_A$ extending to high bias. It reveals strong nonlinearities and a critical current behaviour[26] appearing as peaks in the differential resistance ($dV/dI$). The low-resistance state is described by the Drude conductivity, with a current saturating regime occurring at higher bias. At the critical current ($I_C$), the strong electric field triggers Schwinger-like particle production[26], creating a highly dissipative electron-hole plasma (Fig. 1A).

Interestingly, the nonlinearities are extremely sharp at negative current densities, indicated by the large peaks in $dV/dI$ and discontinuities in the simultaneously measured four-probe DC voltage (Fig. 1C). This is surprising, as it resembles a phase transition, contrasting with previous observations that the evolution is smooth[26]. The origin of this anomalous behaviour is revealed when, instead of using a current source, we switch to biasing our sample using a voltage source. Fig. 1D plots the 2-probe IV characteristics in the vicinity of the transition. Strikingly, at $V_{bias}$ = -1.42 V we observe NDC, when the current starts to decrease upon increasing the bias voltage. NDC extends over a small voltage range after the critical current $I_C$, beyond which the current increases again (with a larger differential resistivity), indicating a transition into the electron-hole plasma phase. This explains why the transition appears sharp in the current-biased scheme (Fig. 1B). The large resistor placed in series imposes a horizontal load line that masks the NDC region and forces the current to jump to the next stable point, creating discontinuities in the IV characteristics (see Supplementary Section 2). Fig. 1E plots the 2-probe IV characteristics when a load resistor of 200 kOhm is placed in series. The voltage jumps abruptly at $I_C$, and hysteresis in the current sweep direction is observed, which is characteristic of a bistable system. The load resistor essentially allows one to engineer a bi-stability from NDC. Fig. 1F plots the differential conductivity d$I$/d$V$ as a function of filling and voltage bias, where NDC (blue regions) appears over a broad parameter space, tracing a ring-like structure characteristic of out-of-equilibrium criticalities in moiré superlattices[26,31]. While there is some asymmetry in the IV characteristics originating from self-gating (see Supplementary Section 3), the NDC appears for both negative and positive bias. This highlights the robust nature of the NDC and its connection to non-equilibrium electron-phases (Fig. 1A), which will be addressed later.

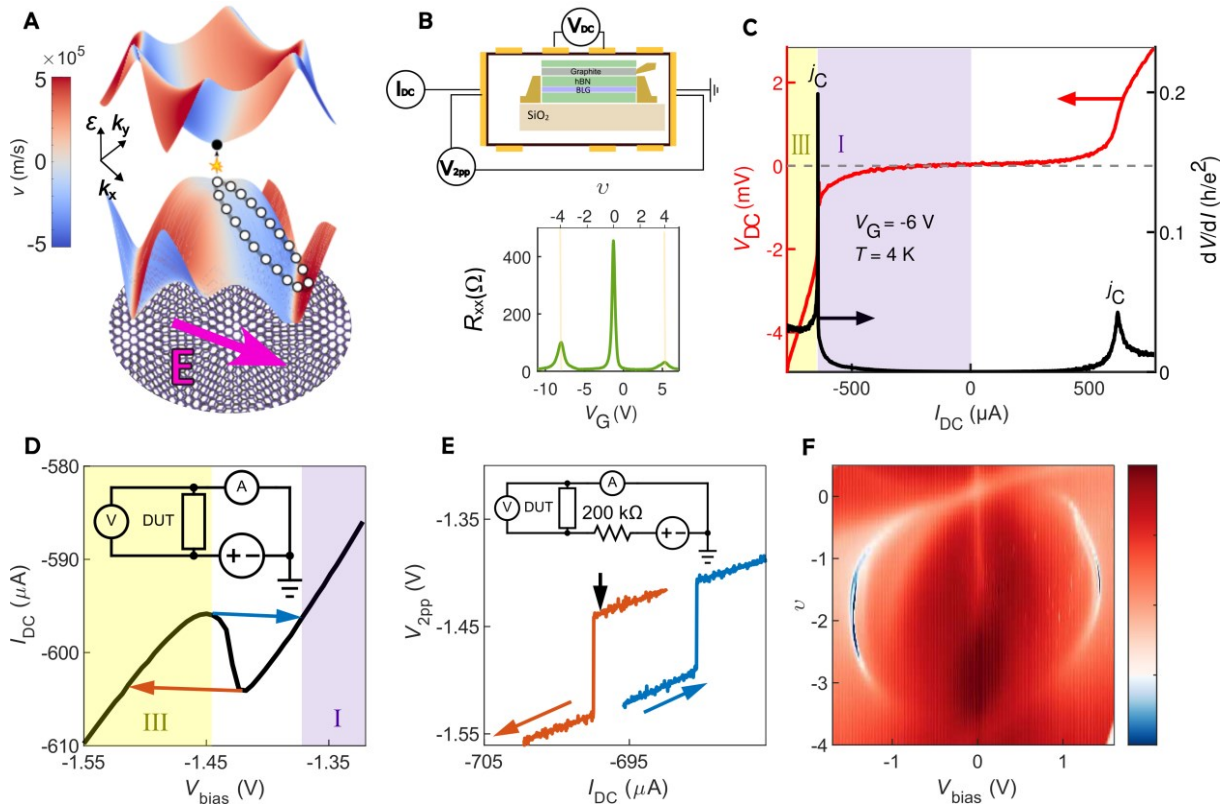

**Figure 1**. **High-field experiments in moiré superlattices**. (**A**) Conduction and valence bands of BLG/hBN superlattice calculated for $D$ = 0.1 V/nm (see Supplementary Section 8), plotted within the superlattice Brillouin zone. Colour shading corresponds to the velocity of holes (open circles) or electrons (filled circles) in the direction of the electric field (pink arrow). With strong electric fields, Schwinger-like transitions are triggered, creating electron-hole pairs. If the gap is sufficiently large to suppress Schwinger-like transitions, carriers may drift into negative-velocity states (red regions). (**B**) Top panel: measurement schematic sketched on a top-down view of the Hall bar mesa. The inset illustrates the heterostructure. Bottom panel: Longitudinal resistance measured as a function of gate voltage ($V_G$) determined from $V_{DC}/I_{DC}$ = 100 nA. The top axis depicts filling factor (ν). (**C**)

Current-voltage characteristics measured in DBLG$_A$. Left axis plots 4-probe voltage ($V_{DC}$, in red) measured at fixed doping around half-filling of the superlattice Brillouin zone ($\upsilon \approx 2$) as a function of current ($I_{DC}$). Right axis plots the corresponding differential resistivity d$V$/d$I$. Inset: circuit schematic. (**D**) Current ($I_{DC}$) plotted as a function of the voltage bias ($V_{bias}$) close to the high-field transition. Inset: circuit schematic. (**E**) 2-point-probe voltage drop ($V_{2pp}$) vs applied current bias, zoomed in on the transition region. Orange and blue solid lines plot forward and backward sweeps, respectively. (**F**) Differential conductivity (d$I_{DC}$/d$V_{bias}$) as a function of filling factor ($\upsilon$) and $V_{bias}$. All measurements were performed at $T$ = 4 K.

**Single-photon detection with field-induced bistabilities**

The jumps in the current-voltage characteristics close to $I_C$ form the core ingredients required for sensitive photodetection. Indeed, when biasing close to $I_C$ (black arrow in Fig. 1E), we find that the system stochastically switches from the metallic to the electron-hole plasma state. This is shown in Fig. 2A, which plots the 2-probe voltage drop recorded over time ($t$), when the system spontaneously jumps to a high resistance state at $t$ = 211 ms. To restore the system, we apply resetting voltage pulses, with reset frequencies reaching up to 38 kHz depending on the measurement conditions. We attribute the observed jumps to the electrical noise and thermal radiation in the system, which demonstrates the extreme sensitivity of the bistability. Led by this result, we record similar time traces under mid-infrared illumination (11.32 µm). Notably, we observe many more switching events (Fig. 2B), demonstrating that light increases the count rate ($\Gamma$). Fig. 2C plots $\Gamma$ as a function of $I_{DC}$ near the transition for different illumination intensities, revealing two fundamental characteristics. First, $\Gamma$ increases exponentially as $I_{DC}$ approaches the transition (see Supplementary Section 16 for a detailed analysis). Second, close to $I_C$, $\Gamma$ increases linearly with power ($P$) and exhibits super-linear behaviour when biased further from the transition (Fig. 2D). Both observations bear strong similarities with the operation of SNSPDs. The former indicates the probabilistic nature of switching events[32,33], while the latter provides the first indication of single-photon detection, where the number of switching events is directly proportional to the photon flux[34]. More generally, the exponent in the $P$ dependence $\Gamma \propto P^x$ is characteristic of the number of photons necessary for triggering a switch and increases further away from the critical bias. This is evidenced in our data by a range of biases showing $x$ = 1, while those further away from the $I_c$ tend towards $x$ = 2 (Fig. 2D).

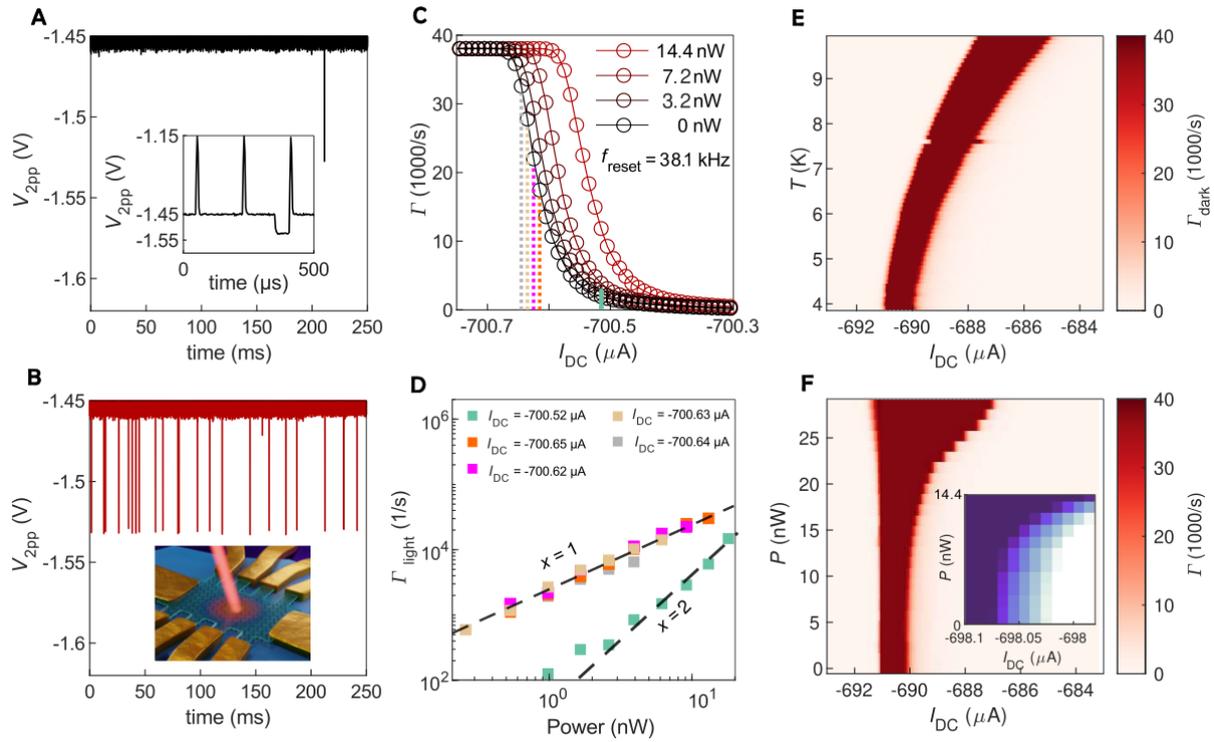

**Figure 2 Single-photon counting at mid-infrared frequencies.** (**A**) 2-probe voltage ($V_{2pp}$) plotted as a function of time (ms) when current-biased close to the critical current $I_C$. The reset frequency $f_{reset}$ = 5.6 kHz. We apply voltage pulses at frequency of $f_{reset}$ = 5.6 kHz to reset the system. Inset: Zoom of main panel displaying three reset pulses and one click. (**B**) Same as (A) but under MIR illumination. $I_{DC}$ = -701.2 μA, λ = 11.32 μm, P = 4.8 nW, $f_{reset}$ = 5.6 kHz, filling factor (ν) = - 2. Inset: schematic of device and photocurrent experiment. (**C**) Count rate (Γ) plotted as a function of bias current ($I_{DC}$) for different powers (P) (black to red). $f_{reset}$ = 38 kHz (**D**) count rate only due to light, $Γ_{light} = Γ - Γ_{dark}$, plotted as a function of P for different $I_{DC}$. Filled squares are experimental data points, with colours corresponding to the bias levels marked with dashed lines in panel (**C**). The dashed lines mark dependencies $Γ \sim P^x$ for x = 1 and 2. (**E**) $Γ(I_{DC},T)$ measured in the dark. (**F**) $Γ(I_{DC},P)$ measured at λ = 11.32 μm. $f_{reset}$ = 38 kHz. Inset: same as (**F**), but measured in a lower-noise configuration, focusing on the low-power regime. We note the doping level in (**E**) and (**F**) was slightly different than measurements in (**C**), which accounts for slight differences in $I_c$ between these measurements. The P reported in all panels refer to absorbed powers and are calculated using rigorous coupled wave analysis method (see Supplementary Section 19) to be 4 %. All measured were performed at 4 K except **E**.

The observation of linear P dependence of Γ (Fig. 2D) points towards the detection of single photons. However, the presence of switching in the dark (Fig. 2A) brings into consideration the effect of light-induced heating. To disentangle the different contributions, we measure the temperature dependence of Γ. Fig. 2E plots Γ as a function of $I_{DC}$ and temperature (T) in the dark, whereas Fig. 2F plots Γ as a function of $I_{DC}$ and P. The dark red regions mark the count rate equal to the reset frequency ($Γ = f_{reset}$= 38.133 kHz), indicating detector saturation. The two maps are markedly different. In Fig. 2E the saturation region shifts to lower current densities with temperature. We attribute this behaviour to a global shift in the critical current as temperature increases, a trend characteristic of the NDC in our samples (see Fig. 5A). On the other hand, the $Γ = f_{reset}$ region does not shift but rather broadens with increasing P. The critical current remains the same but the onset of switching occurs for a bias further from the transition. Moreover, analysis of the $Γ(I_{DC})$ dependencies (Supplementary Section 16) shows they are strongly modified when changing $P$ while they remain unchanged with variations in temperature. A detailed analysis of the temperature dependence of the dark count rate (Supplementary Section 16) reveals further differences. It increases exponentially with temperature exhibiting activated behaviour, in contrast to the linear response observed in the power dependence (Fig. 2D). These differences suggest that mid-IR sensitivity arises from direct photon absorption, rather than global sample heating.

Without using a single-photon emitter, an alternative proof for single-photon detection involves analysing the switching statistics of the detector and correlating it with the emission statistics of a dim coherent source[35]. In the mid-IR, such measurements are challenging in free-space coupling schemes because background radiation dominates our dark counts and mixes with the statistical analysis (see Supplementary Section 5). Therefore, to characterize the underlying mechanism, we performed measurements using visible light (675 nm) where background mid-IR radiation was heavily filtered using a quartz window (see methods). Additionally, the shorter wavelength enables spatially resolved measurements providing further fundamental insights into the detector operation. For spatially resolved measurements, we employ an alternative measurement scheme which yields a quantity $\Gamma_{arb}$ proportional to $\Gamma$ (see Supplementary Section 4). Fig. 3A plots $\Gamma_{arb}$ as a function of position ($X,Y$) of the beam, with the device outline overlaid; the device mesa is mapped out by tracking the zero-bias photocurrent originating from contact regions (see Supplementary Section 4). Curiously, it shows that $\Gamma_{arb}$ is maximum in a narrow strip of the device (see further discussion below). Parking the laser there, we studied the dependence of $\Gamma$ on $I_{DC}$ and laser power $P$ (Fig. 3B), revealing a similar behaviour to the mid-IR case (Fig. 2B); note that count rate is less than mid-IR experiments only because we employed a lower reset frequency for these measurements (5.6 kHz) and does not reflect spectral dependencies. Fixing $I_{DC}$, we perform statistical measurements of the switching events. Coherent sources emit photons according to a Poissonian distribution[36] and observing that distribution reflected in the count statistics is a commonly accepted benchmark of single photon detection[2,5,19,35]. Fig. 3C plots histograms of the counts recorded in 235 ms time bins, demonstrating that they obey Poissonian statistics (pink lines) for finite $P$. Fig. 3D shows that the histograms satisfy the relation variance = mean, characteristic of Poissonian distribution, for the entire dynamic range of the detector. These results demonstrate that our device can indeed detect single photons. While the statistical measurements for mid-IR require more controlled experimental conditions, the stochastic switching behavior, linear power dependence, and exclusion of laser-induced heating (Fig. 2) provide sufficient evidence for the mid-IR capabilities of our detector.

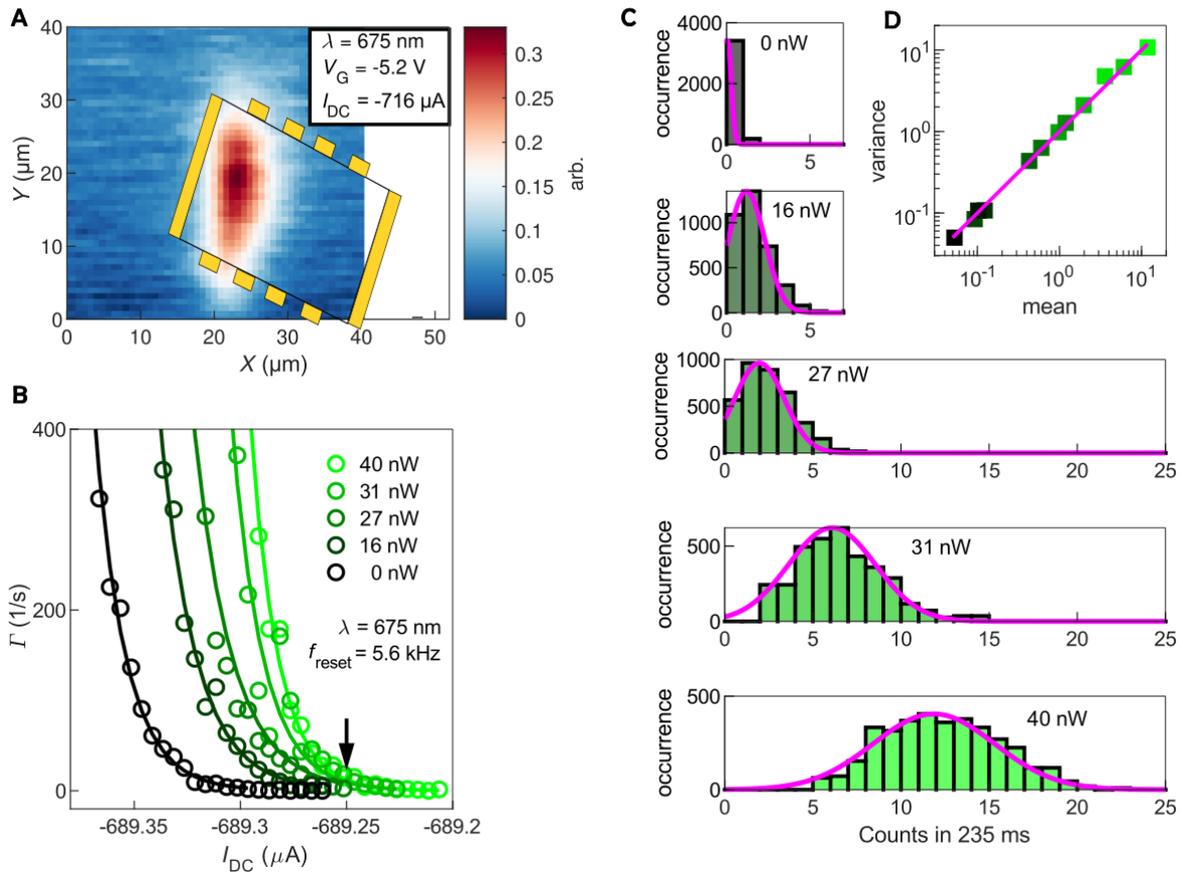

**Figure 3. Spatial and statistical measurements of switching events using visible light.** (**A**) Measurements of $\Gamma_{arb}$ as a function of spatial position (*X,Y*) in the two point probe geometry ($V_{2\text{-probe}}$). The scale bar marks the relative values of $\Gamma_{arb}$ (see Supplementary Section 4 for details on the measurement scheme). The device mesa and measurement contacts are sketched in black and gold, respectively. (**B**) $\Gamma$ as a function of $I_{DC}$ plotted for different powers (*P*) (black to green). The solid lines plot exponential fits. The reset pulse frequency is 5.6 kHz (lower than in Fig. 2) and the plot is truncated at $\Gamma$ = 400 to better see the features at low $\Gamma$ values. The displayed *P* are corrected for BLG absorption (2% at λ = 675 nm, see Supplementary Section 19). (**C**) Histograms presenting the statistical distribution of count number registered in a 235 ms time bin for different illumination *P*, recorded for the bias value marked by black arrow in (B). The distributions follow Poissonian statistics (pink solid lines). (**D**) Variance plotted as a function of mean calculated from the histograms measured for twelve different *P* values color scale corresponds to the main different *P* presented in (B). The solid purple line plots variance=mean line, indicating a Poissonian distribution. All measurements were performed at *T* = 4 K.

### Origin of NDC

The data above provide exciting prospects for single photon detection. The first step towards finding the design principles necessary for the development of MSPDs is understanding the mechanism of NDC. This phenomenon is characteristic of superlattice structures[37] and, broadly, can be split into two classes. First, the regime of sequential tunnelling in weakly coupled superlattices, also known as electric-field localization, which occurs when the voltage drop ($V_{SL}$) per superlattice period (*a*) exceeds the mini-bandwidth[38] (*W*): $V_{SL}/a > W$. The second class refers to the Esaki-Tsu negative differential velocity (NDV) in strongly coupled superlattices[39,40] where carriers are accelerated to negative velocity states at the inflection points of the Brillouin zone (see Supplementary Section 6). While these peculiar quantum effects are not accessible in normal metals, the superlattice imposes a small Brillouin zone (10,000 times smaller than that in regular crystals), making such phenomena readily accessible in moderate electric fields. Electric-field localization is more common in narrow-bandwidth materials[38], in which the condition $V_{SL}/a > W$ is easier to satisfy, whereas NDV pertains to wider bands. Although our moiré system resembles strongly coupled superlattices, disentangling these two mechanisms is challenging owing to the uncertainties in the material parameters and electric field distributions in our devices. Nonetheless, they can be distinguished based on the dependence of NDC on $W$[40]. Specifically, the threshold for electric-field localization increases linearly with $W$, whereas the threshold for NDV is generally independent of $W$. In our moiré heterostructures utilizing bilayer graphene, we can in-situ tune $W$ using displacement fields[41] and hence study the dependence of the critical field ($E_C$) on $W$. For this purpose, we turn to the second device with two electrical gates, the DBLG$_B$.

Because in DBLG$_B$ the NDC is localised to one side of the device (see Supplementary Section 11), it is masked by parallel non-NDC channels. Hence, to track the NDC response, we measure the 4-probe voltage ($V_{4pp}$) in the region where bi-stabilities were observed (see Supplementary Sections 10 and 11). The inset of Fig. 4A plots the $V_{4pp}$ as a function of the source voltage in DBLG$_B$ for different displacement fields ($D$) close to $I_C$. By increasing $D$, we observe that the NDC region becomes more pronounced, but its critical voltage ($V_C$) remains little influenced. At the same time, the measurements of the critical current (see Supplementary Section 10) show a decrease in $I_C$ by 50 % for the largest $D$, providing experimental evidence for reducing $W^2$ (i.e. band flattening) with $D$. Fig. 4A plots $V_C$ as a function of $W$, where with the D-dependent $W$ is obtained from calculations (see Supplementary Section 8). The red dashed line plots the critical voltage expected for electric-field localization. Our experiment clearly does not follow this trend, and the measured $V_C$ is mostly independent of $W(D)$, suggesting that the observed NDC originates from negative differential velocity rather than sequential tunnelling.

Guided by these insights, we performed Boltzmann transport simulations of electron transport in realistic band structures (see Supplementary Section 7) to obtain IV characteristics in the high-field regime. The green curve in Fig. 4B shows the calculated current density as a function of $E$-field at a fixed doping density. It shows that carriers' drift velocity first increases before reaching the critical field, at which it begins to drop. To visualize the microscopic origin of the NDV, we plot the distributions of carriers for zero bias (Fig. 4D, left), at $E_C$ (Fig. 4D, middle) and in the NDC regime (Fig. 4D, right), with arrows representing the contribution of individual states to the total current. At zero bias, the carriers are evenly distributed around the Brillouin zone and the total current is zero. As $E$ increases, the carrier distribution shifts and the current grows until, around $E_C$, more and more carriers start to enter the regions with negative velocity (arrows pointing in the direction opposite to the applied field). Fig. 4C plots the calculated critical current $I_C$ at which NDC onsets, compared with that extracted from our experiment (Fig. 1E) as a function of doping. Without any fitting parameters, the experiment and calculations agree within an order of magnitude. Moreover, the calculations capture the filling dependence with the largest critical current peaking at $\nu = \pm 2$. This is attributed to the presence of van Hove singularities at $\nu = \pm 2$ for which the free carrier density is largest, and the current is maximum. Beyond $E_C$, the green line in Fig. 4B deviates significantly from the experiment. This is unsurprising since the calculations do not consider Schwinger-like transitions[42] to neighbouring bands, which are responsible for the continued growth of the current beyond the NDC regime[26] and were found to dominate in previous high-field transport studies in moiré superlattices without a gap at charge neutrality[40]. However, our BLG/hBN superlattices exhibit a gapped band structure. If sufficiently large, the gap suppresses these transitions, allowing the carriers to continue accelerating within their bands and reach the NDV regime. This explains why in DBLG$_B$, the NDV appears only when gapping the CNP strongly with displacement fields (Fig. 4A inset) and provides evidence for the requirement of a gapped state. The interplay of NDV with Zener transitions may also explain the doping dependence of the NDC regime, appearing only for the highest doping in the middle of the bands. However, a complete understanding requires considering a two-band model that accurately captures Zener transitions.

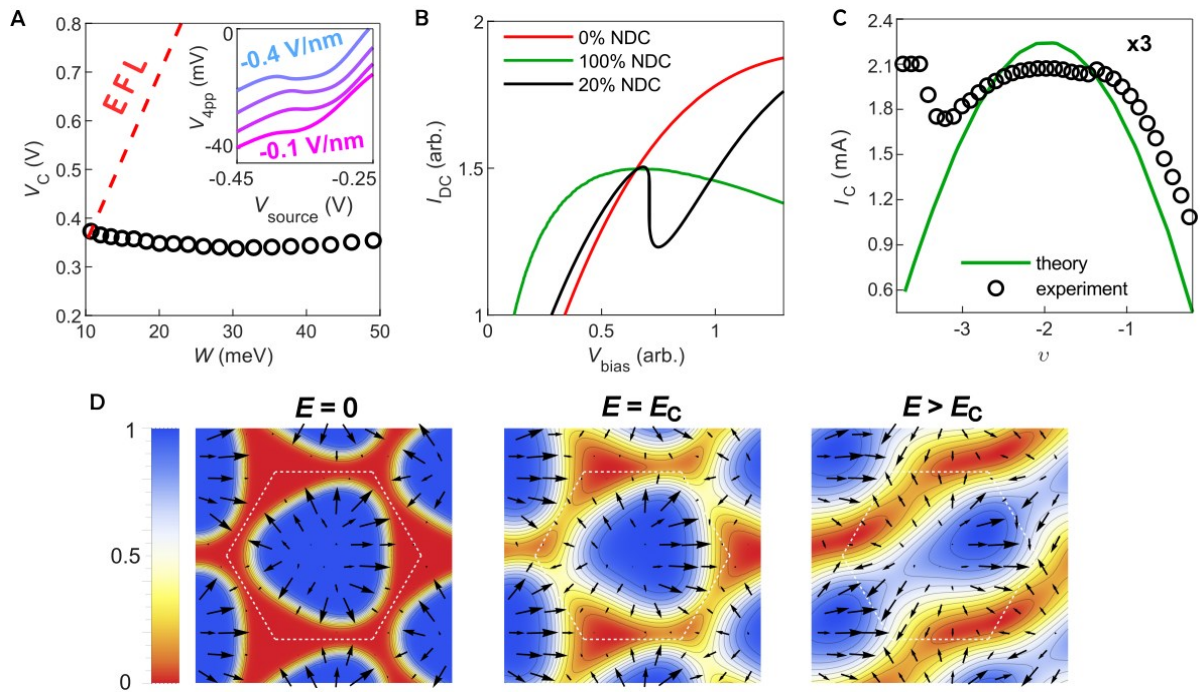

**Figure 4. Displacement field dependence of NDC and modelling negative velocity.** (**A**) Critical voltage ($V_C$) as a function of bandwidth ($W$). Inset: Four-probe voltage $V_{4pp}$ as a function of voltage bias $V_{source}$, measured for different displacement fields ($D$). $V_c$ in the main panel was obtained from analysing this trend (see Supplementary sections 8 and 10 for details on $V_c$ and $W$ extraction). (**B**). The two-region model shows how the NDC (green) and non-NDC (pink) regions connected in series can result in different IV characteristics (black), depending on the relative contribution of the two regions (See Supplementary Section 13 for details). (**C**) Critical current ($I_C$) as a function of doping. Green line is calculated for BLG/hBN superlattice considering the carrier dynamics within the lowest-lying band. Open circles plot experimental data of DBLG$_A$, scaled by a factor 3. Extraction of $I_C$ from experiment is unreliable for $v < -3$ due to low signal (*cf.* Fig. 1F). (**D**) Carrier distributions within a single Brillouin zone (white dashed hexagon) for electric fields below, at, and above critical field $E_C$, obtained from the same model as the green curves in panels **B** and **C**. The colour scale plots probability of occupation and the black arrow direction and magnitude plots the relative direction and amplitude of group velocity of those carriers in the direction of the applied field.

Further understanding of the measured NDC can be obtained from our spatially resolved measurements (Fig. 3A, Supplementary Section 11), which show photoactivity only in a narrow strip of the device, suggesting that the NDC is localized to a small domain within the sample. The local nature of the effect is further confirmed via transport measurements performed between different contact pairs, where NDC can only be observed if the current path crosses this region of the device (see Supplementary Section 12).

We also note that the photoactivity remains fixed to this region for all gate voltages and bias currents/directions (see Supplementary Section 17) suggesting the domain is not field-induced but rather structural in origin. The spatial dependence hence likely reflects local regions of favourable stacking arrangements, which can vary significantly in moiré materials[43,44]. Although the domain in $DBLG_A$ appeared close to current injectors, the count rate was seemingly independent of the polarization of light (Supplementary Section 14) suggesting the effect originates from the bulk rather than metallic contacts. Moreover, similar localized regions observed in $DBLG_B$ were found in the centre of the device, microns away from the current injectors providing stronger evidence for domain structures in the bulk of the device (Supplementary Section 11).

While the lowest-lying bands are easily gapped with displacement fields, they remain mixed with remote bands for most stacking configurations, and hence Schwinger transitions are still expected to dominate. However, there exist a few stackings resulting in a gap between the lowest-lying and remote bands (see Supplementary Section 8). Statistically, they are far less common but may be found in the large devices used in this study. Indeed, transport measurements in the linear response regime showed varying behaviours between different contact pairs, including signatures of supermoiré lattices[45,46], providing strong evidence for spatial variations in stacking (see Supplementary Section 1). This explanation also clarifies why our NDC is seemingly tied to the Schwinger criticalities (Fig. 1F). A simple circuit model (See Supplementary Section 13) demonstrates how the regions without this stacking-induced gap, where Schwinger criticalities dominate (pink curve in Fig. 4B), influence the IV characteristics of the entire device when connected in-series with the NDC region. This results in a compression of the NDC region and the combined IV characteristics (black curve in Fig. 4B) that resemble the experimentally measured data (Fig. 1D). Consequently, the measured NDC appears only for a narrow bias range as a result of interplay between the two regions.

**Single-photon detection mechanism**

Our detector uniquely integrates engineered light-sensitive bistabilities with the broadband mid-infrared absorption inherent to bilayer graphene.[47] The switching mechanism shares similarities with both Geiger-mode operation in resonant tunnelling diodes[48], and SNSPDs, where the system transitions between conductive electron phases (Fig. 1C). Phenomenologically, the switching is triggered by a photon-induced current or voltage, which drives the system into the unstable negative differential conductance (NDC) region, initiating the switch. Given that our electron system is metallic, photo-excited carriers likely perturb the local current distribution, leading to the formation of a localized hot-spot with a reduced critical current. This hot-spot expands, eventually exceeding the threshold for switching. This interpretation is supported by the strong temperature dependence of the globally measured critical current (Fig. 2E).

The underlying microscopic process remains an open question, particularly in understanding how a single photon generates a sufficiently large photocurrent to trigger the global switching event. This typically requires some form of gain or carrier multiplication. Photoconductive gain may play a role in amplifying the current generated by a single-photon, if electron-hole lifetimes invoke the asymmetries intrinsic to the hexagonal-boron nitride/graphene superlattice[30]. On the other hand, our findings indicate that the photoactive region is spatially confined, possibly exhibiting high-field domains which may enable carrier multiplication via impact ionization[49], assisting hot-spot formation. Further experimental work is required to fully elucidate the microscopic processes governing single-photon switching in this system.

**Discussion**

The distinct underlying physics of our MSPDs means they are not constrained to the same operating conditions as SNSPDs. Their operating temperatures are instead limited by the robustness of NDC and the accompanying bi-stabilities. Fig. 5A plots IV characteristics for increasing temperatures showing the NDC persists up to around 30 K. Fig. 5B plots $\Gamma$ as a function of $\Gamma_{dark}$ for temperatures up to 25 K and under mid-IR illumination, where larger $\Gamma_{dark}$ indicates larger bias currents closer to the transition. For all $T$ we find light-induced switching with little deterioration in detector performance up to 20 K. Fig. 5 demonstrates the higher $T$ capabilities of our MSPDs, which may be pushed to even higher $T$ through engineering the superlattice minibands.

In conclusion, we developed a new platform for single-photon detection based on non-equilibrium electron phases in moiré superlattices with tuneable band structures. In our un-optimized devices and measurement protocols, we demonstrate a unique combination of detection capabilities including single-photon sensitivity extending to mid-IR wavelengths (>10 µm), fast response (> 38 kHz, setup limited), high temperature operation (25 K), and the potential for achieving high detection efficiencies. While our demonstrated detector operation was limited to the probabilistic regime[32,33], with inherently low efficiencies, the exponential dependence of the count-rate on bias current suggests high detector efficiencies are attainable if dark counts are reduced and the reset rate can be increased. To achieve this, fibre-coupled device packages may be engineered with sharp band pass filters to supress background photons[50]. Similarly, the 38 kHz speed and the detection range we report represents lower bounds, limited by our experimental set-up (see Supplementary Section 15).

Now that the proof-of-concept device has been demonstrated, there are a myriad of future directions to explore. At the material level, engineering the NDV and non-equilibrium electron phases through moiré heterostructure engineering is crucial to achieving high performance. This could lower critical currents and accompanying shot noise, expand the photoactive region, and potentially increase sensitivity for longer wavelengths. In this regard, other moiré systems or van der Waals superlattice structures may also be explored, and may even be better suited for observing NDC. For example, we also observed high-field induced bi-stabilities in magic-angle twisted bilayer graphene[51,52] (See Supplementary Section 18) showing that the NDC is commonplace to different moiré materials. At the device level, the geometry of the mesa can also be improved significantly. Large current injectors help to minimize joule heating effects which may smear the NDC, while narrower channels can lower the critical current. The latter is analogous to the behaviour of SNSPD's that employ nanowire structure for enhancing internal quantum efficiency. From the measurement side, improving optical coupling and devising automatic re-setting schemes will enable much faster and sensitive operation, approaching and potentially surpassing those of SNSPDs which are limited by kinetic inductance.

Our MSPDs pave the way for a novel class of compact, broadband and CMOS-compatible SPDs, with potential applications ranging from astrophysics to LiDAR, molecular spectroscopy and quantum technologies[3,53]. While our detection mechanism is unique to the high-field properties of moiré superlattices, we note that alternative NDC-based devices may also be explored, combining the capability to engineer sensitive bi-stabilities with resonances at mid-infrared wavelengths. Given bilayer graphene's intrinsically large long-wavelength response, our devices may even have potential for photodetection in the terahertz regime[30,47]. Simultaneously, the manifestation of NDC opens new avenues for high-field experiments in moiré heterostructures. Our observation of NDV represents a significant milestone towards realizing Bloch oscillations in two-dimensional solid-state systems, where several intriguing phenomena[54,] and potential device technologies reside, including high-frequency oscillators[55], multi-valued logic[56], and compact THz emitters.

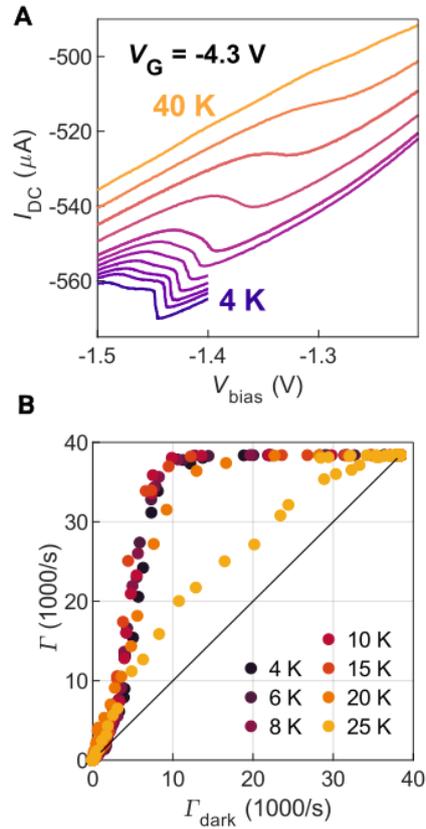

**Fig. 5 High-temperature performance.** (**A**) IV curves showing NDC as a function of temperature. The NDC region shifts to lower bias and flattens with increasing the temperature and the NDC survives up to around 30 K. (**B**) Correlation plot relating $\Gamma$ to $\Gamma_{dark}$ for a few different temperatures. Every trace is in fact a bias sweep but this way of plotting the data allows for a direct comparison between different temperatures, as changing the temperature moves the optimal operating bias point. The detector performance stays unchanged up to around 20 K.

**Methods**

**Device Fabrication:** The samples presented in this work are fabricated using the typical van der Waals heterostructures assembly technique. Typically, a thin hBN flake (~10-15 nm) with at least one straight edge is picked using the hot-pick-up technique[57], with a polypropylene carbonate (PC) film on a polydimethylsiloxane (PDMS) stamp at 90º C. This hBN flake is then later used to pick up a Bernal BLG flake, mechanically exfoliated on $Si^{++}/SiO_2$ (285 nm) from highly oriented pyrolytic graphite, and pre-characterized using optical microscopy, and Raman spectroscopy[58]. The alignment is done such that the straight edge of hBN is parallel to the straight edge of BLG resulting in 50% probability of forming an interfacial moiré superlattice. hBN has a honeycomb lattice structure like graphene but with a ~ 1.8 % larger lattice constant due to a longer B-N bond than C-C. This lattice mismatch results in a moiré period up to 14 nm when aligned perfectly[59], thereby creating a moiré superlattice. Finally, the stack is used to pick up a last layer of hBN and later dropped down on a pre-patterned marker chip of Si$^{++}$/SiO$_2$ (285 nm) at 180°C, squeezing out the bubbles, and impurities as previously reported. The stack is then shaped into a Hall bar geometry using SF$_6$ plasma, and O$_2$ plasma to etch top hBN, and BLG respectively[57], and further metalized using 3/15/30 nm of Cr/Pd/Au. Lastly, to independently control both charge carrier density and electric field inside the BLG a top gate is fabricated using a few layers graphene encapsulated in hBN, patterned and contacted similar to the bottom stack. The top gate stack is shaped such that all the edges of BLG are uniformly gated to open a clean bandgap in the BLG.

In DBLG$_A$, two independent hexagonal-boron nitride layers were selected for the encapsulation, and aligned to the bilayer graphene pristine crystallographic edge. The resulting heterostructure had close alignment of both top and bottom hBN and, consequently, certain regions of the device show signatures of supermoiré potentials (Supplementary Section 1). In DBLG$_B$, the bottom hBN was aligned while the top one was rotated by 30°. This ensures that one and only one of the hBN layers is guaranteed to have alignment to the crystallographic axis. In DBLG$_A$ the bottom gate was leaking and hence only the top gate could be tuned to control carrier density and displacement field. In DBLG$_B$ both gate electrodes were functioning, allowing independent control of *n* and *D*.

**Cryogenic and electrical measurements:** All our measurements were performed in a Montana s50 Cryostat with variable temperature between 4-300 K and optical access. For the linear response, electrical measurements were performed using lock-in amplifiers (SR860) at frequencies 13 -27 Hz sourcing a constant voltages of $V_{AC}$ ~500 µV. For high-bias measurements and electrical gating, Keithley 2400 source meters were used. Finally, DC voltage signals were measured using ITHACO voltage preamplifiers 1201 and read out digitally using a digitization acquisition card DAC (NI 6363).

**Mid-Infrared and visible wavelength photocurrent measurements:** For mid-infrared photocurrent measurements, we use a Daylight MIRcat tuneable QCL laser operating at wavelength λ = 11.32 µm. The power is limited by an ND filter and precisely controlled by a set of ZnSe holographic wire grid polarizers from Thorlabs. The polarization of the outgoing beam is kept fixed throughout the experiment by the last polarizer. The light is focused by a reflective objective with Numerical Aperture (NA) = 0.5 and 13 mm aperture. The beam is expanded by a set of parabolic mirrors to match this aperture. For visible wavelength measurements we employed a Fabry-Perot diode (λ = 675 nm, model S1FC675 from Thorlabs). The objective is a 20x Mitutoyo Plan Apo Infinity Corrected Long WD Objective with NA = 0.42 and WD = 200 mm. The beam is collimated, cleaned with a pinhole and expanded to assure a uniform gaussian profile with the beam diameter closely matching the objective aperture. The power was controlled by two polarizers (LPVISC-100 from Thorlabs), always keeping the resulting beam polarization fixed. In both measurements, the focal point is scanned by moving the objective with a motorized XYZ stage. The displacement of the optical axis of the objective is negligible on the scale of beam diameter. For mid-IR measurements a ZnSe window (1 mm thickness) was used with transparency 2 – 20 µm (Edmund Optics) while for the visible range fused silica windows (3 mm thickness) with transmission 200 nm – 2 µm were used. This change of window was the only modification to the experimental set-up. Therefore, we attribute the differing dark counts between the two wavelength regimes to the background radiation coming from our laboratory which is transmitted by the ZnSe window and partially blocked by the fused silica window.

**Click-counting measurement scheme:** For single-photon counting measurements an AC resetting pulse with a square waveform and 4% duty cycle is generated by a waveform generator (Agilent

33600A). The DC current is measured with a DAC (NI 6363) after preamplification with ITHACO current preamplifier 1211. The DC voltage drop over the sample is measured using a DAC (NI 6363).

The count rate ($\Gamma$) is obtained from a recorded waveform according to the following procedure. First, we differentiate the waveform. By differentiating the waveform, we avoid the influence of any standing wave in our circuit affecting our click counting procedure. Then we define the click threshold at 0.02 V below the median of the whole waveform. Any point that falls below 0.02 V from the median triggers counting a click.

**Acknowledgments**

We thank Ilya Charaev for numerous, in-depth technical advices and reviewing the manuscript. We also thank Giacomo Scalari, Jérôme Faist and Leonid Levitov for helpful discussions. PJH acknowledges support by AFOSR grant FA9550-21-1-0319, the Gordon and Betty Moore Foundation's EPiQS Initiative through Grant GBMF9463, the Ramon Areces Foundation, and the ICFO Distinguished Visiting Professor program. F.H.L.K. acknowledges support from the ERC TOPONANOP (726001), the government of Spain (PID2019-106875GB-I00; Severo Ochoa CEX2019-000910-S [MCIN/ AEI/10.13039/501100011033], PCI2021-122020-2A funded by MCIN/AEI/ 10.13039/501100011033), the "European Union NextGenerationEU/PRTR (PRTR-C17.I1), Fundació Cellex, Fundació Mir-Puig, and Generalitat de Catalunya (CERCA, AGAUR, 2021 SGR 01443). Furthermore, the research leading to these results has received funding from the European Union's Horizon 2020 under grant agreement no. 881603 (Graphene flagship Core3) and 820378 (Quantum flagship). This material is based upon work supported by the Air Force Office of Scientific Research under award number FA8655-23-1-7047. Any opinions, findings, and conclusions or recommendations expressed in this material are those of the author(s) and do not necessarily reflect the views of the United States Air Force. J.B. acknowledges support from the European Union's Horizon Europe program under grant agreement 101105218. R.S., B.J., M.V.M. and L.C. acknowledge support from Research Foundation-Flanders (FWO) research projects G0A5921N and 11E5821N. H.A., K.N. and R.B. acknowledge funding from the European Union's Horizon 2020 research and innovation programme under Marie Skłodowska-Curie grant agreement no. 665884, 713729 and 847517, respectively. V.F. and S.S acknowledge support from EPSRC Grants EP/S019367/1, EP/P026850/1, and EP/N010345/1; British Council project 1185409051. D.B.R. acknowledges funding from the Secretaria d'Universitats i Recerca del Departament d'Empresa i Coneixement de la Generalitat de Catalunya, as well as the European Social Fund (L'FSE inverteix en el teu futur)-FEDER. R. K. K acknowledges funding by MCIN/AEI/ 10.13039/501100011033 and by the "European Union NextGenerationEU/PRTR" PCI2021-122020-2A within the FLAG-ERA grant [PhotoTBG], by ICFO, RWTH Aachen and ETHZ/Department of Physics, and support from the Ramon y Cajal Grant RYC2022-036118-I funded by MICIU/AEI/10.13039/501100011033 and by "ESF+"

## Supplementary Information

**S1 Quantum transport characterization of bilayer graphene aligned to hBN heterostructures**

In this section, we present quantum transport measurements performed in our bilayer graphene (BLG) – hexagonal boron nitride (hBN) superlattice devices, whose optical images are presented in Fig. S1A/B for $DBLG_A/DBLG_B$, with the contacts labelled. The panels below the optical images plot the 4-point probe resistance ($R_{xx}$) measured between different contact pairs for a fixed AC voltage (500 µV RMS) sourced through the injector probes. In general, the transport characteristics show behaviour consistent with the literature, with a large peak in resistance at charge neutrality and satellite peaks which indicates full filling of the first superlattice miniband. The satellite peaks have a much lower resistance than the charge neutrality point (CNP), suggesting that the first and second mini-bands are not gapped, in agreement with previous calculations[1,2]. However, for certain contact pairs in $DBLG_A$, namely contacts B-C and G-H, one can find additional fine features which are not expected for the case of bilayer graphene aligned to a single hBN layer. Indeed, the edges of the top and bottom hBN flakes (white lines in Fig. S1A) are nearly parallel to each other, suggesting that bilayer graphene may be aligned to both hBN layers. The additional peaks in resistance between some pairs of contacts may reflect a supermoiré structure[3,4] originating from the beating of two bilayer graphene/hBN moiré potentials. These marked differences in quantum transport between different contact pairs illustrate how the electronic properties are varying locally in our devices. These structural variations likely originate from changes in stacking order and twist angle between different regions[5,6,7]. which can influence also the non-linear electrical response, and explains why the negative differential conductance (NDC) may be localised to a specific region in our devices (Fig. 3A, Fig. S11).

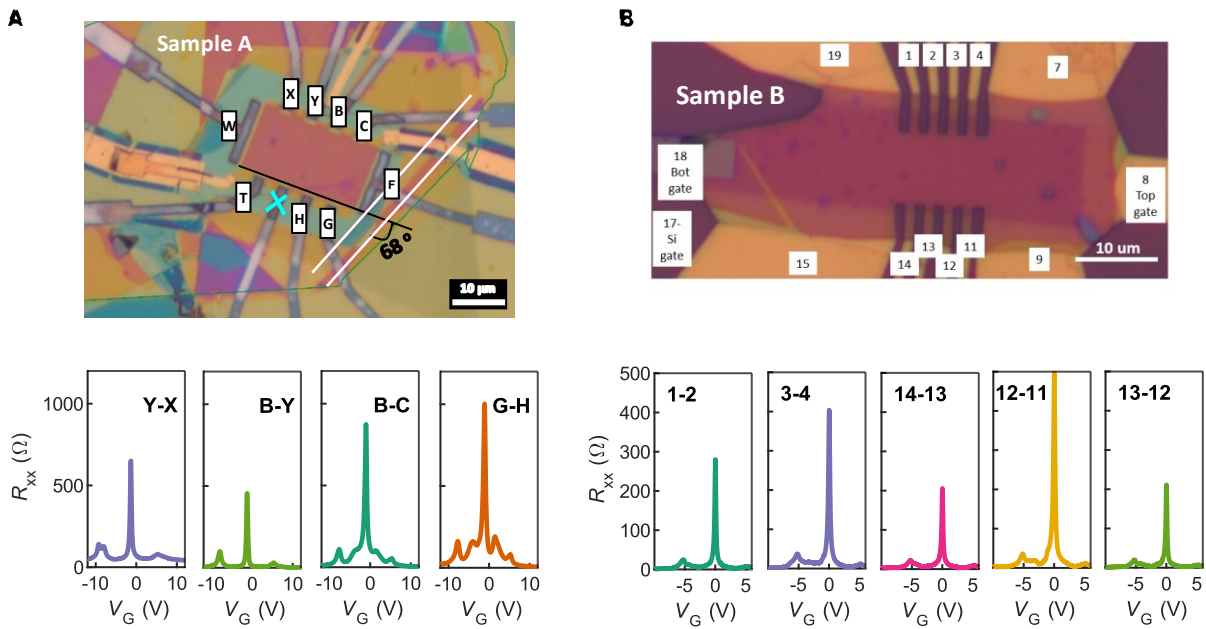

**Fig. S1 – homogeneity of the devices. A)** Top panel – Optical image of the sample (50 x magnification). The contacts are labelled with capital letters. The edges of hBN flakes are marked by white lines and the edge of BLG flake is marked with a green line. The edge of the Hall bar (black line) makes an angle of approximately 68° with the edge of hBN flakes. Contact crossed with teal X is not functioning. Bottom panels plot Resistance ($R_{xx}$)

measured as a function of the top gate voltage ($V_G$) measured between different contact pairs. Current was sourced between contacts W and F. **B)** Top panel – Optical image of the sample B with contact labelled numerically (50x magnification). Bottom panels – $R_{xx}(V_G)$ between different contact pairs. AC voltage was sourced simultaneously to contacts 19 and 15, while the current was drained from contacts 7 and 9. For both samples the bottom gate voltage was set to 0 V, with the driving voltage of 0.5 mV RMS amplitude at 13.1 Hz frequency.

## S2 Engineering bi-stabilities with load resistors

In non-linear circuits that include NDC elements, the IV characteristics can become unstable when combined with other components. Placing a load resistance in series with an NDC element limits the stable operating points in the circuit. Fig. S2A plots the load line (blue dashed line) on top of the intrinsic current-voltage (IV) characteristics of our device (black) a with a 2 kOhm resistor placed in series. One can see that the load line appears rather vertical in comparison to the IV characteristics of our device and hence there is only one stable operating point for all the applied voltages. The resulting circuit can operate still in the NDC regime except with a slightly less pronounced slope. However, placing a 200 Kohm resistor in series creates multiple operating points for a fixed source voltage (Fig. S2C). In this case, the operating points with negative slopes are unstable and the system instead jumps to the stable points with positive differential resistance. The two stable operating points define the bistability in the circuit. However, if the operating voltage separating the two stable points is large enough, then the operating point with minimum voltage change (most energetically favorable) is selected. Hence, the system follows a stable IV characteristic until it reaches the onset of NDC where it jumps to the next stable point. Depending on the sweep direction, the system jumps from point 2 to 3 (forward) or 4 to 1 (backward) tracing a hysteresis (Fig. 2C). Shining light causes the system to jump between the stable states, specifically from points 2 to 3. We note that the 200 kOhm series resistor imposes a nearly horizontal line (Fig S2B) and increasing the resistance further has little effect on the system. Decreasing the resistance, however, will make the load line steeper, effectively narrowing the hysteresis loop. Tuning the bistability and accompanying hysteresis may allow for enhanced performance including, for example, lower voltage resetting pulses.

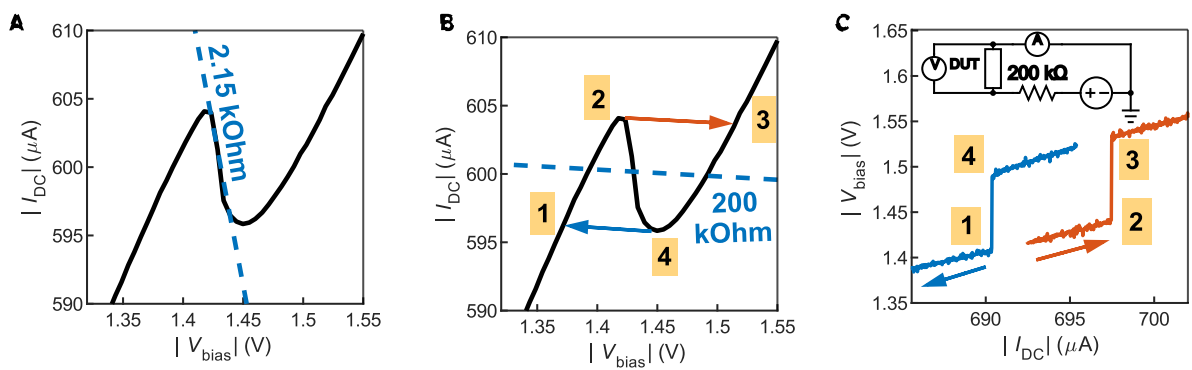

**Fig. S2 – Engineering bi-stability. A)** *IV* curve measured by sourcing voltage with the load line for a 2.15 kOhm resistor drawn over for $V_B$ = -2.63 V. **B)** The same *IV* curve with the load line for 200 kOhm for $V_B$ = -121.3 V. The numbers correspond to the points marked in panel c. **C)** Hysteresis loop in the *VI* curve measured by sourcing current, using a voltage source and a 200 kOhm resistor placed in series (see inset). The operating points 1-4 are marked as in (B)

## S3 Self-gating contributions to the IV characteristics

Self-gating is a phenomenon in-which the in-plane voltage bias becomes comparable to the gate voltage applied to the device. In this case the in-plane voltage bias not only drives the current but also influences the free carrier density in the channel. For low bias voltages in the linear response regime such effects are typically neglected. However, under high bias voltages such phenomena start to manifest, leading to peculiar non-linear responses. For example, the asymmetry in $I_C$ reported around Fig. 1B originates from this. In this section we provide additional measurements where we correct for self-gating effects and demonstrate that the NDC still persists, distinguishing its origin.

Figure. S3A plots the current ($I_{DC}$) as a function of the gate voltage ($V_G$) for different voltage biases ($V_{Bias}$). Depending on the sign of $V_{Bias}$ the data shows a maximum in the current corresponding to the gate voltage when the system is doped at the charge neutrality point (CNP). Increasing from negative to positive $V_{Bias}$ we notice this maximum shifts monotonically towards $V_G = 0$. This shift corresponds to the self-gating where the bias off-sets the effective gate voltage. Fig. S3B plots the gate voltage where CNP occurs for different bias voltages which follows a quadratic-like behaviour. To correct for this, we simultaneously sweep the gate voltage and bias voltage to maintain the same carrier density at each voltage bias. The correction is evident in Fig. S3C which shows how all bias voltages exhibit the maximum in $I_{DC}$ for the same $V_G$. Fig. S4D plots dI/dV similar to Fig. 1F of the main text but using the self-gating correction. The ring-like structure becomes symmetric and the NDC remains confirming it does not originate from self-gating effects like those reported previously in single layer graphene devices.[8]

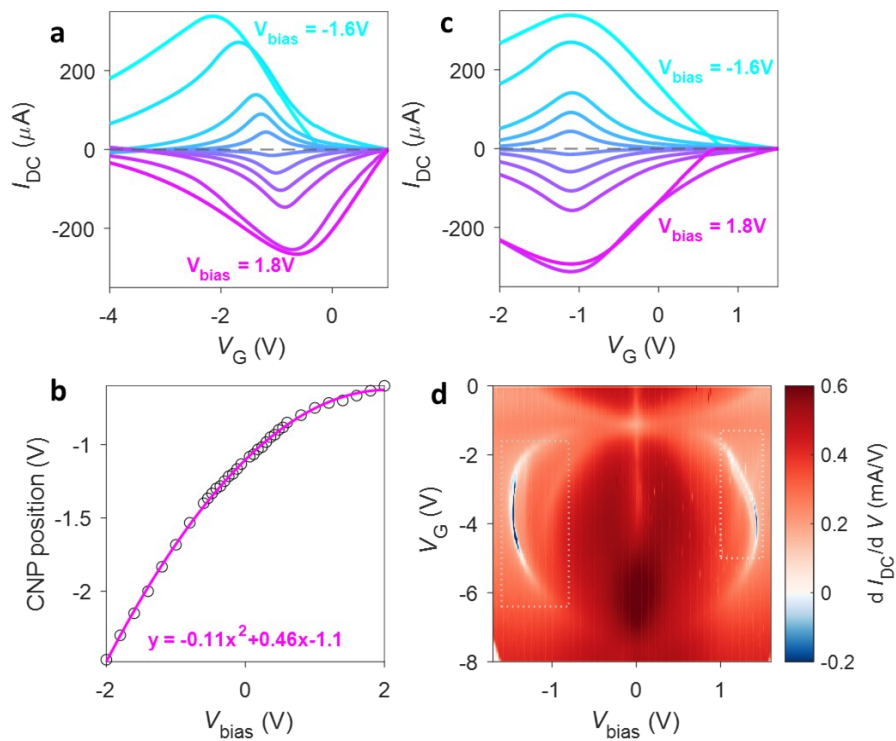

**Figure. S3 A**, current ($I_{DC}$) measured as a function of gate voltage ($V_G$) in DBLG$_A$. The cyan to blue datasets correspond to different voltage bias $V_{bias}$ between contacts W and F. **B**, the position of charge neutrality point (CNP) extracted from **A**, plotted as a function of $V_{bias}$. Open circles are experimental data and the solid magenta line is a quadratic fit. **C,** $I_{DC}(V_G)$ plotted for similar $V_{bias}$ but correcting for self-gating. **d,** differential conductivity $dI_{DC}/dV_{bias}$ as a function of $V_G$ and $V_{bias}$ measured correcting for self-gating (**C**).

## S4 Spatial characterisation measurements of single photon counting

To characterize the spatial dependence of the single-photon counting (Fig. 3A of main text), we first mapped out our sample using scanning photocurrent microscopy under zero-bias. In the absence of applied bias, photocurrent measurements performed in our devices are sensitive to photoactive junctions that form close to contacts and generate a photocurrent via the photo thermoelectric effect[9]. Fig. S4A plots a scanning photocurrent measurement performed under zero bias between two contacts indicated on the figure. Photoexcitation is achieved using red light with a spatial resolution of around 1 μm. While there is some spatial dependence throughout the device area, most of the signal is concentrated at two hot spots of opposite sign where a bilayer graphene – gold contact forms a photoactive PN junction. Changing the contact configuration (Fig. S4B), we find hot spots in different regions of the spatial map indicating the position of those new junctions formed by the contacts. Performing these measurements between different contact pairs allows us to build a spatial map of the contacts on the device and the underlying Hall bar mesa (drawn schematically on Fig. S4A/B).

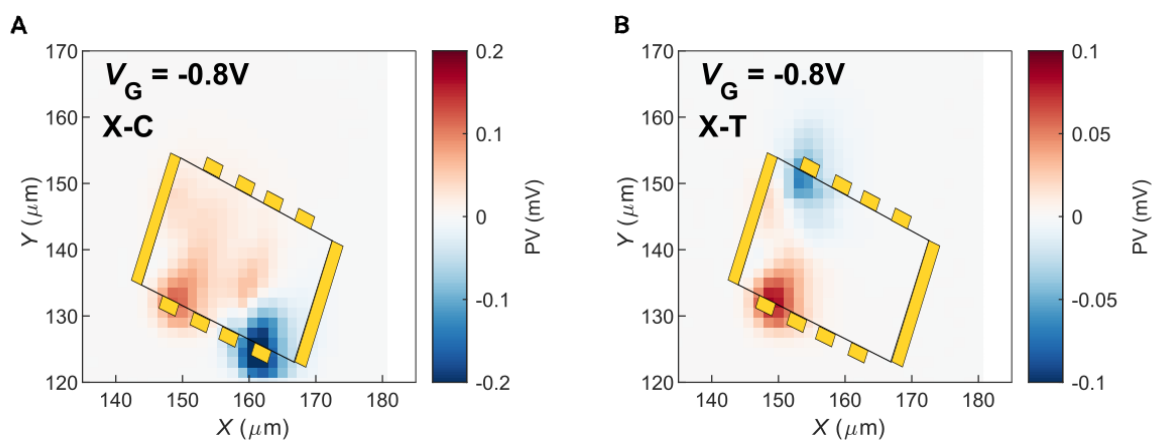

**Fig. S4 – Determining the sample orientation.** Exemplary maps of the photovoltage generated under red light illumination (λ = 675 nm), in which we can locate the contact pair X-C (panel A) and X-T (panel B). The contact X, which should appear at the same position in both maps, is highlighted in red, allowing for identification of contacts C and T. Repeating this procedure with a few other contact pairs for increased precision allows for an accurate determination of the sample.

In Fig. 3A of the main text, we presented spatial measurements of the photon counting. Measuring the count rate requires long measurements of time traces and significant post-processing to obtain the statistical values. This makes it difficult to do high-resolution spatial mapping where many data points must be collected in a short time to avoid drift. Therefore, we employed a different method for measuring the photon counting events. The circuit used is shown in Fig. S5. The DC voltage is sourced from the Keithley source meter (Keithley 2400) through a large resistor (200kOhm) to operate voltage/currents close to the bistability. After registering a click, the system becomes stuck in the electron-hole plasma state. Therefore, in order to bring it back to the 'active' state we must apply a resetting current pulse. To this end, an AC sine wave of ~6.2 μA RMS amplitude is generated from a lock-in amplifier (SR860) through a 100 kOhm resistor at a frequency of 870.1 Hz. The amplitude is large enough to reset the system after a click. Because the resistance of the system increases after a click event, the mean magnitude of the AC voltage drop measured in the sample is higher for a higher count rate and shows as an increased signal on the lock-in amplifier readout. Hence, the voltage drop measured in this way reflects the photon count rate we refer to as $\Gamma_{arb}$, and provides a fast method to perform spatial dependent measurements. Although the exact relationship between the measured voltage and the count rate is unknown, the method allows to assess qualitative changes such as the spatial dependence.

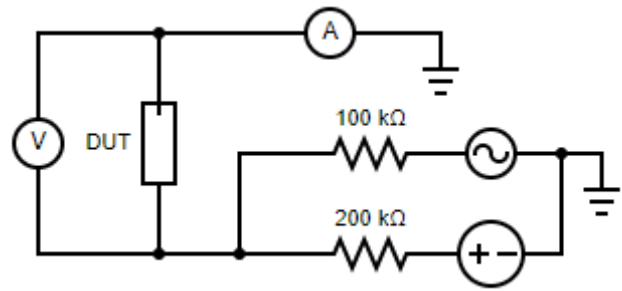

**Fig. S5 – Circuit for count rate mapping.** The circuit used to measure spatial maps of the count rate with red light. DC current is sourced from a voltage sourcemeter (Keithley 2400) through a large resistor (200kOhm) placed in series with the device. The AC sine wave of ~6.2 µA RMS amplitude is generated from a lock-in amplifier (SR860) at a frequency of 870.1 Hz. The AC voltage drop over the sample (DUT) is measured with the lock-in amplifier (marked as a voltmeter in the schematics).

**S5 Statistical Measurements of count rate in the Mid-infrared**

As shown in Fig. 3 of the main text, the photon counting statistics for visible wavelengths followed a Poissonian distribution, proving strong evidence for single photon switching. However, for the mid-IR there is a notably larger contribution from dark counts. In this regime, the switching statistics deviate significantly from Poissonian. We attribute this behaviour to a dominating contribution from background mid-IR photons to the dark counts.

Background radiation remains a challenge in long-wavelength single photon detection measurements, requiring sophisticated methods. Our wavelength of interest (11.3 um) lies very close to the peak of black body radiation at room temperature (peaks at around 10 um) making isolation especially challenging. For our MID-IR measurement, the sample is placed in a cryostat with an optical access, equipped with two ZnSe windows with transparency window between 2- 20 um (Fig. S6a) and with mid-IR radiation coupled from a quantum cascade laser (QCL) through a reflective objective. To decrease the influx of stray photons hitting the chip whether coming from the lab or the outer case of the cryostat, we have covered the sample with thick aluminum foil, thermally anchored to the cold finger of the cryostat and kept at the temperature of 4 K; a small (~4mm diameter) hole was made in the foil, to let the laser beam through. This already significantly decreased the stray photon contribution allowing better detector sensitivity. Nonetheless, we found background photons still perturbed our measurements. This is evident in Fig. S6B which plots the dark count rate $\Gamma$ for three scenarios, without a window (steel blank), with the window (no – blank) and with the window but with tissue paper placed on the outside. It shows clearly that dark counts increase by orders of magnitude when the window is used indicating background radiation from the lab is responsible for the dark counts. Even with the steel cover, however, one does not get rid of stray photons. The cover itself, as well as the outer ZnSe window, are kept at room temperature still emit black body radiation that reaches the detector. The effect of this radiation is hard to evaluate.

Fig. S6C/D plots histograms of the counts recorded in the dark and under mid-IR illumination. Because of these dominating contributions to the dark counts, the distribution is non-Poissonian. Future works may use fibre-coupled systems[10] or employ bandpass filters[11] to lower dark counts induced by background radiation, as well as lower the electrical noise to enable proper statistical analysis in the mid-IR range.

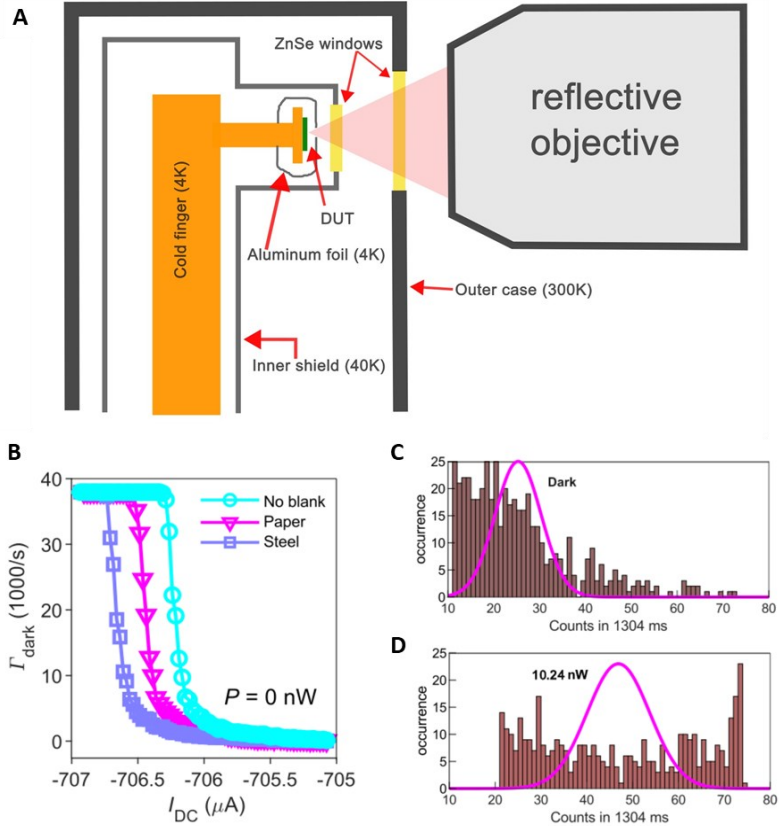

**Fig. S6 – Mid-IR set-up, Dark counts and non-Poissonian distributions. A)** Schematic of the set-up used for performing single-photon detection measurements in the mid-IR wavelength range **B)** The dark count rate as a function of bias current for three different stray lightning conditions. Steel blank completely blocks the stray illumination, paper blank partially transmits light and no blank stands for the ZnSe window of the cryostat without any extra filtering. Different traces were measured in a random order, showing the reproducibility of the measurement. **C)** Dark counts histogram recorded for DCR = 10000 at $f_{reset}$ = 38.1 kHz, corresponding to the onset of linear power dependence in Fig. 2d. The distribution is clearly non-Poissonian (the best fit of Poissonian distribution is outlined in pink). **D)** The same experiment performed under illumination with $P$ = 1.28 nW and $\lambda$ = 11.3 μm. At such a low power most of the counts still come from dark counts, so the resulting histogram is a convolution of photon counts (with unknown distribution) with dark counts that are highly non-Poissonian.

### S6 Esaki-Tsu model - Bloch oscillations and van Hove singularities (1D)

In Supplementary Section 7, we describe a simple generalization of Esaki - Tsu [12,13] approach to the case of a 2D degenerate fermion system which was used to calculate the results presented in Fig. 4C of the main text. Beforehand however, it is instructive to consider the original one-dimensional model to understand the basic carrier dynamics of electrons accelerated to inflection points in the Brillouin zone. According to the semi-classical Boltzmann transport theory, the constant applied electric field ($E$) affects the electron crystal momentum $k$ following[14]:

$$\hbar \frac{dk}{dt} = Ee \, ,$$

where ℏ is reduced Planck's constant, $t$ is time and $e$ is electron charge. After integration this yields

$$k(t) = \frac{Eet}{\hbar} + k_0, \quad (1)$$

Where $k_0$ is the carrier's crystal momentum at $t = 0$. Knowing $k(t)$ we can find the instantaneous velocity $v(t)$ of the carrier in a band which is defined:

$$v(t) = \frac{1}{\hbar}\frac{d\varepsilon}{dk(t)}, \quad (2)$$

Where $\varepsilon$ is energy. The average velocity of the single carrier for a fixed *E*, *i.e.* the single carrier drift velocity $v'_d$, is then given by the time integral of $v(t)$ multiplied by the probability $p(t)$ that the carrier does not scatter within the time $t'$:

$$v'_d = \frac{1}{\tau}\int_0^\infty p(t')\,v(t')dt'. \quad (3)$$

where $p(t')$ is the probability the particle with lose momentum after a time $t'$. A simple approximation for $p(t')$ is the relaxation time approximation, $p(t) = \exp(-t/\tau)$, where $\tau$ is the average time between scattering events, the scattering time.

We analyze the electron dynamics in for two distinct band dispersions, the original sinusoidal dispersion proposed by Esaki and Tsu[12,13] given by,

$$\varepsilon(k) = \frac{\Delta}{2}(1 - \cos(kd)), \quad (4)$$

and a second one, which approximates a specific direction in within the Brillouin zone of bilayer graphene aligned to hexagonal-boron nitride by the following analytical expression:

$$\varepsilon(k) = \frac{\Delta}{2}(1 - 0.5\cos(kd) - 0.5\cos(2kd)),$$

The two dispersions are plotted together in Fig. S6A/S6B, together with the corresponding carrier velocity and carrier effective mass $m^*(k) = \frac{d^2\varepsilon}{dk^2}$.

Let's consider a single charge carrier living in this band. The effect of *E* is to change the velocity of the carrier (eq. 1, 2) with the rate depending on the field magnitude, which in the case of a sinusoid band (eq. 4) is captured by:

$$v(t) = v'_0\big(\sin(\omega_B t + k_0 d)\big), \quad (5)$$

where $v'_0 = \frac{\Delta d}{2\hbar}$ and $\omega_B = eEd/\hbar$. As shown in Fig. S6A, the carrier velocity $v(k)$ peaks near the middle of the Brillouin zone where the dispersion is steepest, and then steadily drops to zero at the edge of the Brillouin zone. Accordingly, the effective mass $m^*(k)$ crosses zero where the velocity $v(k)$ peaks and has a local minimum at the Brillouin zone edge. Evidently, after the point at which $m^*(k) = 0$, the effect of DC electric is to *decrease* the carrier velocity $v(k)$. As a result, after a sufficiently long time, $v(t)$ becomes negative, and for sufficiently strong fields, may even oscillate between positive and negative values. This is commonly referred to as a Bloch oscillations[15,16] and is illustrated in Fig. S6C. It plots the evolution of carrier velocity in time for two values of electric field, $E = 0.2E_c$ and $E = E_c$, where $E_c = \hbar/ed\tau$ corresponds to the electric field required for carriers to gain crystal momentum equal to the Brillouin zone lattice vector in the time of one $\tau$. Evidently, in the absence of scattering, the carrier velocity oscillates sinusoidally in time. However, in the presence of momentum-relaxing scattering, this oscillation will be damped in time. Hence, it is the scattering that causes the average carrier velocity in time, *i.e.* the drift velocity $v'_d$, to be nonzero. After carrying out the integral in eq. (3) (with τ assumed to be energy-independent), $v'_d$ is found to be

$$v'_d(k) = v'_0 \left( \frac{\sin(k_0 d) + \frac{E}{E_c}\cos(k_0 d)}{1 + \left(\frac{E}{E_c}\right)^2} \right), \tag{6}$$

and can be plotted as a function of *E* (Fig. 6c). Initially, $v'_d$ rises linearly with $E$, in accordance with the Ohm's law. For higher fields the relation becomes nonlinear and $v'_d$ reaches a local maximum at $E = E_c$, after which it starts to *decrease* with *increasing E.* This is the onset of negative differential velocity (NDV). Consequently, the current ($j = nev'_d$) starts to decrease with increasing *E*; the phenomena known as negative differential conductance (NDC). Any material that exhibits zeros in the $m^*(k)$ may undergo such behavior, provided that τ is sufficiently long for carriers to reach these parts of the Brillouin zone. In naturally occurring crystals, this is usually not the case because the Brillouin zone is so large that $E_c = \hbar/ed\tau$ attains enormous values, well over the breakdown field value in common materials. However, in superlattice structures like our moiré superlattices, the Brillouin zone is significantly smaller, lowering $E_c$ to experimentally feasible values.

The situation is analogous in the double-sinusoid dispersion (Fig. S6B), with the difference that now $v(k)$ has two local minima and two local maxima. In this case, the effective $m^*(k)$ crosses zero relatively close to the Brillouin zone center, causing a sign change in $v(k)$ well before reaching the Brillouin zone edge. According to eq. (2), $v(t)$ is

$$v(t) = \frac{v'_0}{2\big(\sin(\omega_B t + k_0 d) + 2\sin(2\omega_B t + 2k_0 d)\big)}, \tag{7}$$

which yields

$$v'_d(k) = v'_0 \left( 0.5 \frac{\sin(k_0 d) + \frac{E}{E_c}\cos(k_0 d)}{1 + \left(\frac{E}{E_c}\right)^2} + \frac{\sin(2k_0 d) + 2\frac{E}{E_c}\cos(2k_0 d)}{1 + \left(2\frac{E}{E_c}\right)^2} \right), \tag{8}$$

both of which are plotted in Fig. S6D and Fig. S6F, respectively. The van Hove singularity occurring in the middle of the Brillouin zone causes the local maximum in the drift velocity to occur at lower electric field compared to having a van Hove singularity at the edge of the Brillouin zone only. The full period of Bloch oscillation is not affected. This observation is relevant for realistic cases including bilayer graphene aligned to hBN, where this behavior occurs only for a specific direction.

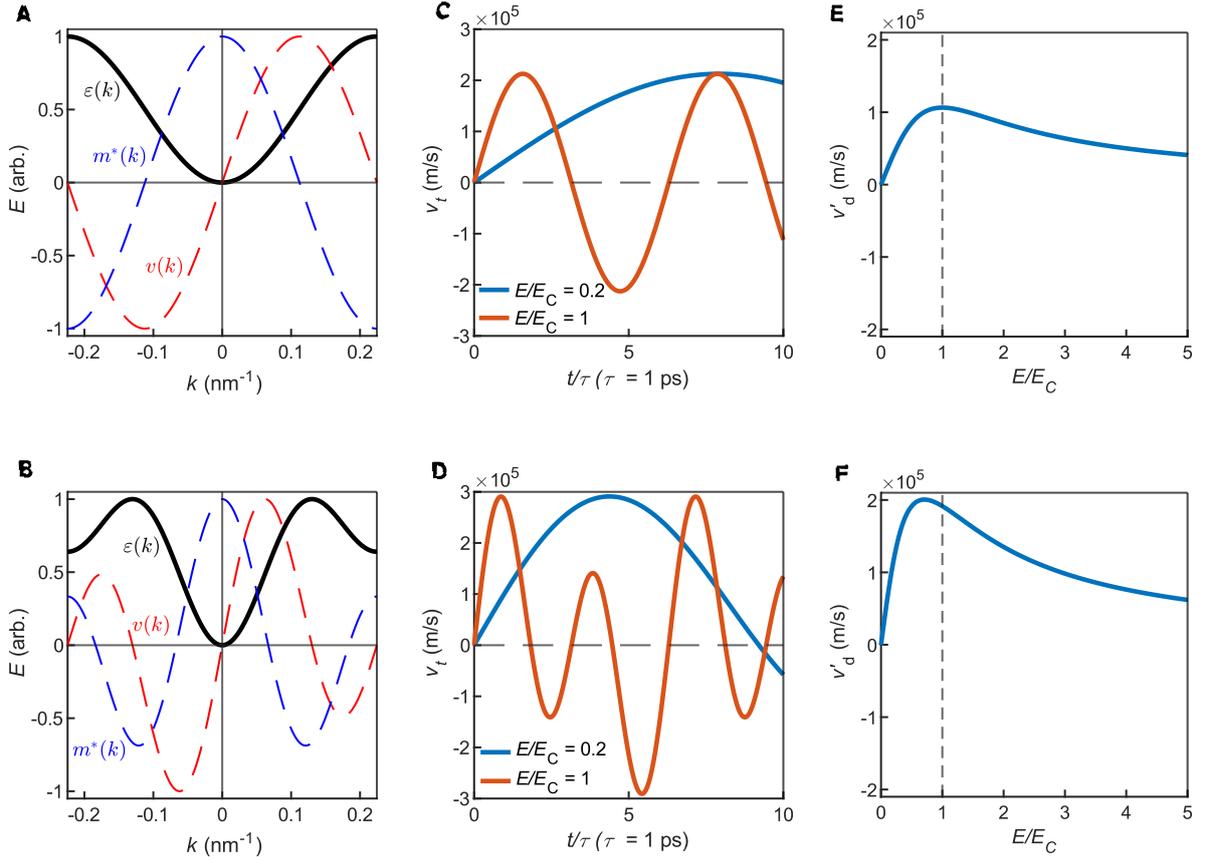

**Fig. S6 Esaki-Tsu model. A)** The sinusoidal dispersion $\varepsilon(k)$ with corresponding velocity $v(k)$ and effective mass $m^*(k)$, where all these quantities are normalized to 1. **B)** Same as panel a but for the double sinusoid dispersion. **C)** Carrier velocity versus time for sinusoidal dispersion at two different amplitudes of electric field, $E$. **D)** Same as panel (C) but for the double sinusoid dispersion. **E)** The drift velocity as a function of electric field $E$ for sinusoidal dispersion. **F)** Same as panel e but for the double sinusoid dispersion. $E_c = \hbar/ed\tau$ is set to the same value in panels (B), (C), (D) and (F).

### S7 Generalized Esaki-Tsu model to two-dimensions (2D)

The calculations of S6 represent the 1D case for electron dynamics in a single band. However, a more advanced treatment is needed to fully describe the two-dimensional case corresponding to our bilayer graphene aligned to hBN superlattice. For this, we generalized the Esaki-Tsu model to the case of 2D degenerate fermion systems including the influence of electron-electron scattering. Our calculations are done using the modelled band structures of bilayer graphene aligned to hBN (see Supplementary Section 8) adopting a numerical approach.

To this end, we formulate and solve numerically the kinetic equation for a single band of 2D electrons in a periodic moiré potential and electric field:

$$\frac{\partial f}{\partial t} + e\, \boldsymbol{E} \cdot \boldsymbol{\nabla}_p f = St[f],$$

where we assume spatial homogeneity but retain the time dependence as a tool for numerical solution of the equation. The collision term is given by:

$$St[f](\boldsymbol{p}) =$$

$$\int_{BZ} f(\boldsymbol{p}')[1-f(\boldsymbol{p})]\, s(\boldsymbol{p}'-\boldsymbol{p}, \epsilon(\boldsymbol{p}')-\epsilon(\boldsymbol{p})) - f(\boldsymbol{p})[1-f(\boldsymbol{p}')]\, s(\boldsymbol{p}-\boldsymbol{p}', \epsilon(\boldsymbol{p})-\epsilon(\boldsymbol{p}'))\, \frac{d\boldsymbol{p}'}{(2\pi)^2}$$

Where $\boldsymbol{p}$ is a vector describing the two-dimensional electron momentum and $f$ is the distribution function in the momentum space. The collision term is modelled as a simple cooling scattering process on an uncorrelated external particle such as impurity, electron or phonon and modelled through a scattering frequency function (analogous to the inverse lifetime), $s(\boldsymbol{p}_i - \boldsymbol{p}_f, \epsilon_i - \epsilon_f)$, which is responsible for both cooling and resistance. For simplicity, we model the scattering frequency as phonon scattering, assuming (due to multiple superlattice folding of phonon branches) that phonon modes exist for any energy above the minimal threshold of $\hbar v_s |\boldsymbol{p}_i - \boldsymbol{p}_f|$ where the sound velocity $v_s \approx 20$ km/s. Hence, we write

$$s(\boldsymbol{p}_i - \boldsymbol{p}_f, \epsilon_i - \epsilon_f) \sim \begin{cases} n_B(\epsilon_f - \epsilon_i, T) & , \epsilon_f > \epsilon_i + \hbar v_s |\boldsymbol{p}_i - \boldsymbol{p}_f| \\ n_B(\epsilon_i - \epsilon_f, T) + 1, & \epsilon_i > \epsilon_f + \hbar v_s |\boldsymbol{p}_i - \boldsymbol{p}_f| \\ 0 & , \text{otherwise} \end{cases}$$

where $n_B$ is a Bose distribution and $T$ is the phonon bath temperature. The solution is found by discretizing the Brillouin zone on a 50 x 50 grid and discretizing time in steps inversely proportional to the electric field, $\delta t \propto 1/|\boldsymbol{E}|$ so that at each time step the distribution function is translated in the direction of applied electric field by a fixed small amount and then cooled by applying the $St[f]$ integrated over $\delta t$. The time evolution is iterated until a stationary state is reached. Examples of distribution function for different applied electric fields are presented in Fig. 4c of the main text. The current is found as $\boldsymbol{j} = e \int \frac{d^2 p}{(2\pi)^2} \nabla \epsilon(p) f(p)$ allowing us to find I-V characteristics up to an unknown constant in the electric field strength. The qualitative form of the result does not strongly depend on the chosen form of scattering.

It is interesting to note that even with an almost ideal cooling assumed in our model, the carriers fail to reach a state resembling a degenerate Fermi gas near the negative differential conductivity regime and beyond. This happens because the high-energy carriers passing over the ridge of the dispersion (See Fig 1B of the main text) at Brillouin zone boundary are no longer scattered back by cooling processes to join the degenerate majority. The classical kinetic description in plane-wave basis used here is expected to break down at very strong electric field (not achieved in the current experiment due to Zener tunnelling effects). In the latter case, the basis of local Wannier-Stark states becomes more appropriate for the description.

**S8 Model of the BLG/hBN superlattice band structure**

The negative velocity effects described above, describe only a single-band model. In real crystals, higher order bands can also influence the non-equilibrium electron phases. If the neighboring bands are spaced closely in energy, carriers may tunnel between them before getting accelerated into the negative velocity states. The phenomena, referred to as Zener/Schwinger transitions, compete with the observation of negative velocity and Bloch oscillations. In previous high-field studies in graphene superlattices, these Schwinger transitions were observed to dominate. In our bilayer/hBN superlattices, we can suppress the Schwinger transitions by opening band gaps in bilayer graphene with displacement fields, which allows negative velocity effects to manifest and results in negative differential conductance. However, transport measurements suggest the bands are not completely isolated, and that the first and second-minibands may even be touching. In this section we perform band structure calculations for different stacking orders and conditions to understand if isolated bands can be obtained in bilayer aligned to hBN, which would explain the dominance of negative velocity effects. Considering that the non-linear response depends strongly on displacement fields (see Supplementary Section 10), it is clear that understanding the details of the band structure is important.

### S8.a Modelling

The band structures are calculated using a large-scale tight-binding approach, in which the whole moiré unit-cell is generated in real space, and the hopping terms are determined using Slater-Koster type functions[17]. Specifically, both intra- and interlayer hopping terms are calculated using:

$$-g(\mathbf{R}) = V_{pp\pi}\left[1 - \left(\frac{\mathbf{R}\cdot\mathbf{e}_z}{R}\right)^2\right] + V_{pp\sigma}\left(\frac{\mathbf{R}\cdot\mathbf{e}_z}{R}\right)^2, \qquad (1)$$

Where:

$$V_{pp\pi} = V_{pp\pi}^0 \exp\left(-\frac{R-a_0}{r_0}\right), \qquad (2)$$

$$V_{pp\sigma} = V_{pp\sigma}^0 \exp\left(-\frac{R-d_0}{r_0}\right). \qquad (3)$$

Here $r_0$ = 0.026 nm is the decay length, $a_0$ = 0.142 nm the interatomic distance in graphene, $d_0$ = 0.335 nm the interlayer distance between graphene layers, $V_{pp\sigma}^0 = 0.48$ eV and $V_{pp\pi}^0 = -2.7$ eV. Next nearest neighbour (nnn) hopping terms $\gamma_{nnn}$ are also included and are calculated using Eq. (1), the decay of Eq. (1) is such that $\gamma_{nnn}$ is 1/10 of the nearest neighbour hopping value. In our tight-binding model we further introduce on-site potentials on the diagonal of the Hamiltonian depending on the atom type, being 0 eV for carbon, 3.34 eV for boron, and -1.40 eV for nitrogen. The values of the parameters are identical as presented in ref.[18]. Relaxation effects are included by relaxing all layers simultaneously using an MD simulation, and recalculating the hopping terms with new atomic coordinates in Eq. (1). A gate voltage in these simulations is applied by adding on-site energy to the corresponding atomic sites in accordance with the real-space function $V(z) = zV_0$, where $z$ is the out-of-plane coordinate. Setting up and solving the Hamiltonian was done using Pybinding[19].

### S8. b, Bandstructures for different stacking orders

The bandstructure of BLG/hBN superlattices for different stacking orders and alignment to hBN layers was studied in detail in ref (9). Fig. S7 plots the band structure calculated for a particular stacking order AB1. In general, similar qualitative dispersions are obtained for the different stackings' orders. Key features include dispersive parabolic conduction and valence bands for the lowest energies, which are separated by gaps for particular configurations and with lattice relaxation. However, for all configurations, displacement fields separate the two lowest lying bands (Figs. S7A-D), producing pronounced band gaps which can reach 50-100 meV, consistent with our measurements. In contrast, the behaviour at the secondary Dirac point (sDP) are quite different, showing a gap at some stacking arrangements but not for others. The presence of a gap is required for supressing Schwinger transitions and allowing negative velocity effects to manifest. Fig. S7E plots the dependence of the bandgap at the SPD on the displacement field $D$. Up to $D$ = 0.3 V/nm the gap size stays fairly unchanged, after which it starts to close due to the flat-band moving closer to the remote band. This result shows that obtaining isolated bands in BLG/hBN superlattices is possible. However, the gap at sDP appears for only a few stacking orders, which makes it statistcally less likely to be observed.

Aside from the gaps being tunable with stacking orders and displacement fields, the bandwidth of the flat-bands is also strongly influenced. The bandwidth of the valence band is plotted as a function of displacement feld in Fig. S7B. Clearly, the bands becomes significantly flatter as *D* is increased, supporting our claim that the critical field's indifference to displacement field means the non-linearities are independent on the bandwidth (*cf.* Fig. 4B of the main text) and provides a strong evidence for the Esaki-Tsu mechanism rather than sequential resonant tunnelling[20].

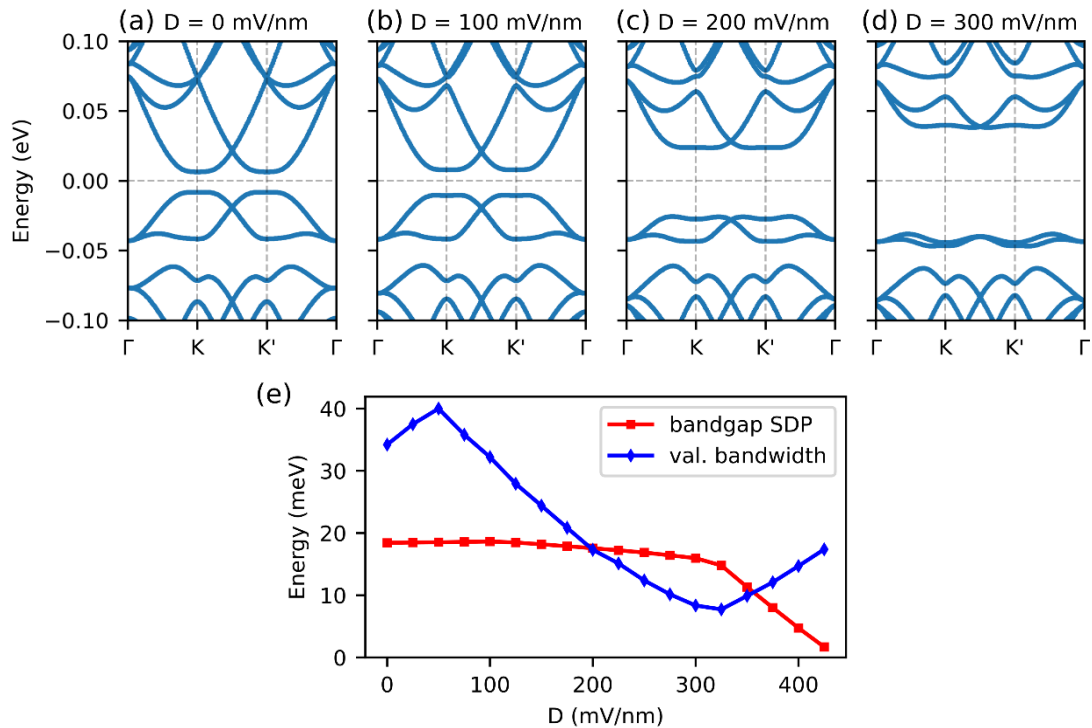

**Fig. S7 Bandstructure calculations.** The calculated dependence of A) the bandgap at the Secondary Dirac point and B) the bandwidth *W* on the displacement field *D*.

### S9 Dependence of critical field on the in-plane field direction

An interesting consequence of a 2D dispersion is that, in contrast to the 1D case, the high-field properties can vary depending on what angle (θ) does the applied in-plane field make with respect to the material's crystal axis. From the optical image of DBLG$_A$ we infer that the source-drain direction is tilted at θ = 8 degrees from the crystallographic axis of the BLG flake. Basing on the simulations and bandstructures introduced in Supplementary sections 7 and 8 we have studied the dependence of high-field IV characteristics on the in-plane field direction. The results are plotted in Fig. S8, showing that the critical field (understood as an onset of NDC) is weakly dependent on the field's angle, except if θ is close to 30 degrees, in which case the critical field is significantly lower. This is an interesting insight from practical perspective, since lowering the critical bias is desirable for device applications and facilitating experimental studies alike.

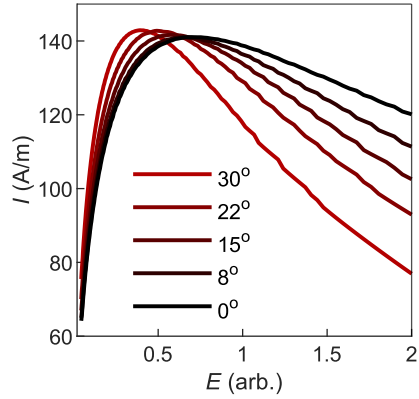

**Fig. S8 Field direction dependence** The IV characteristics calculated in BLG/hBN superlattice band structure for different angles of in-plane electric field (with respect to the crystallographic axis of bilayer graphene). For angles θ close to 30 degrees the critical field is significantly lower.

### S10 Displacement field and doping dependence of the critical current and critical field in device B

As shown in the Supplementary Section 8, the band structure of BLG-hBN superlattice can be tuned in-situ by applying displacement fields (*D*). The *D*-dependent bandwidth and bandgaps, as well as the gate-induced carrier density *n*, are expected to influence the non-linear electrical response in these systems. Here we show these dependencies in sample B where we can tune *n* and *D* separately. Fig. S9B plots the differential conductance (d*I*/d*V*) as a function of filling factor (υ) and voltage bias ($V_{bias}$). Similar to device A, sharp peaks in d*I*/d*V* mark the non-linear response, with a doping dependent threshold voltage tracing a ring-like structure. In contrast to Device A, however, no negative differential conductance was observed in the two point-probe measurements. Fig. S9B plots the two-probe IV characteristics for different displacement fields which introduces gaps between the lowest lying bands. The corresponding differential conductivity (d*I*/d*V*) is plotted in Fig. 9C. It shows the non-linearities are significantly influenced by displacement field. The critical current decreases with displacement field, and the IV curve flattens, with the differential conductivity dipping strongly, almost reaching zero. It does not however become negative, as was the case in sample A. In the simultaneous 4-probe measurement of local voltage drop $V_{3-4}$ as a function $V_{bias}$ (Figs. S10 D,E,F), signatures of negative differential resistance emerged: a small voltage range over which the 4-probe voltage decreases as the bias is increased. We attribute this behaviour to the onset of negative differential conductance occurring locally in a small region between the probe contacts, causing the local 4-probe voltage to drop, whilst the overall differential conductance through the device remains positive. In device B, the bistabilities and hysteresis could be engineered and spatially resolved confirming that the NDC is localised to a small region (Supplementary Section 11 for more details). This analysis tells us that the negative differential resistance (NDR) observed in the 4-probe voltage measurements serve as a signature of local NDC. We elaborate further on that conclusion in Supplementary Sections 11, 12 and 13.

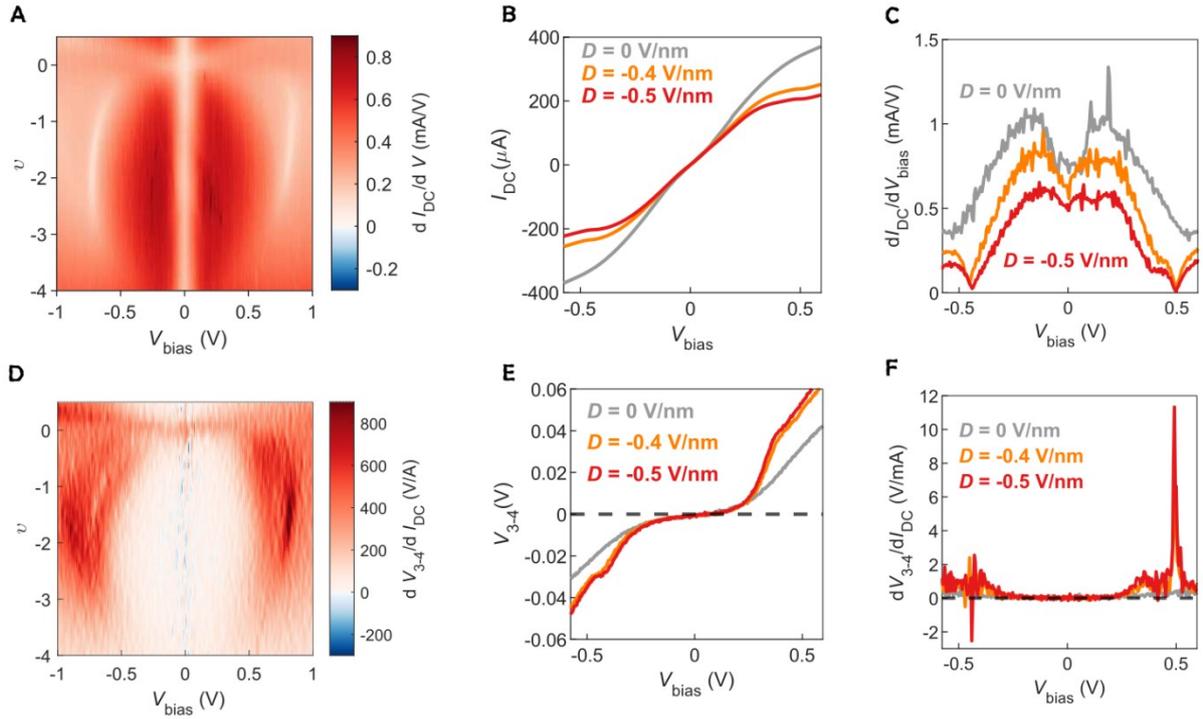

**Fig. S9 High bias measurement in device B. A)** The experimental two-point probe d$I_{DC}$/d$V_{bias}$ map vs source-drain bias $V_{bias}$ and filling $u$, taken with the contact pair 15 - 9 used as the source and contact pair 7-9 used as drain at $D$ = 0 V/nm. In contrast to sample A, NDC is not apparent. **B)** Two-point probe current $I_{DC}$ vs bias voltage $V_{bias}$ measured at carrier density $n$ = -1.25 x $10^{12}$ cm$^{-2}$ for a three values of displacement field, $D$ = 0 V/nm, $D$ = 0.4 V/nm and $D$ = 0.5 V/nm. **C)** The corresponding derivative d$I_{DC}$/d$V_{bias}$ showing how the dip, identified as the onset of negative velocity regime in the local region, shifts with the displacement field. **D)** The experimental four probe differential resistance d$V$/d$I$ vs source-drain bias $V_{bias}$ and filling $u$ measured between probes 3 and 4 simultaneously with panel (A). **E)** The voltage drop between contacts 3 and 4 as a function of the applied 2-probe bias, measured at carrier density $n$ = -0.81 x $10^{12}$ cm$^{-2}$ for different displacement fields. Zoom-in on the plateau at negative biases is shown in Fig. 4 of the main text. **F)** The derivative of panel (E); negative values indicate negative differential resistance (NDR).

**Supplementary Section 11 – Single photon counting and spatial mapping of NDC in device B**

In Supplementary Section 10 we showed how a local negative differential conductance manifests in DBLG$_B$ as a negative differential resistance measured in the 4-probe voltage measurements. Figs. S2A and S2B show how bi-stabilities and hysteresis in different contact pairs can be engineered placing load resistors in series with our device, allowing for performing count rate measurements according to the same procedure as for the DBLG$_A$, and yielding the same result of increased count rate by photoexcitation. This can be seen in Figs. S10C and D, showing time traces in the dark and under MIR illumination, respectively. The bias and power dependence of Γ (Fig S10E) also follows the same behaviour to what was observed for DBLG$_A$ in Fig. 3F of the main text. Moreover, the photoactive region is also concentrated to a small area in the device. Fig. S10F plots spatial measurements of the count rate performed using the measurement scheme described in Supplementary Section 4. However, contrary to device A, the domain does not extend over the whole width of the Hall bar but rather appears centred closer to one side of it. We conclude that the lack of NDC in 2-probe measurement is because the NDC region does not span the whole width of the Hall bar, and is essentially shunted by parallel non-NDC channels. This explains why it can only be seen through the local 4-point probe voltage measurement, while the global 2-point probe current measurement is dominated by the non-NDC regions, showing a dip in d$I$/d$V$ characteristic of the Schwinger criticalities but does not become negative.

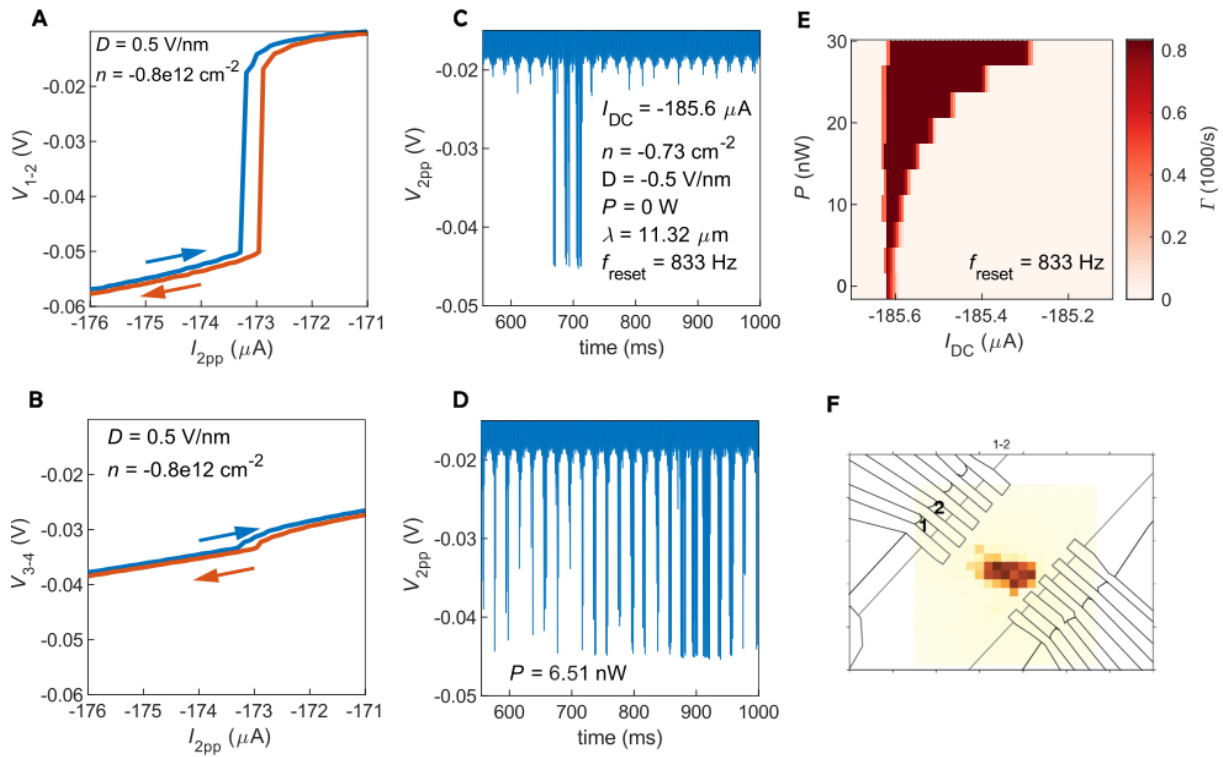

**Fig. S10 Single photon detection in DBLG$_B$** **(A-B)** Hysteresis in *VI* curves measured when sourcing current to the system through a 0.5 MOhm resistor connected in series. Four-point probe voltage measured using contacts 1-2 (A) and 3-4 (B) shows hysteresis arising from the underlying NDC region. C) $V_{2\text{-probe}}$ vs. time recorded without illumination, at $I_{DC}$ = -185.6 μA, $n$ = -0.73 cm$^{-2}$, $D$ = -0.5 V/nm and a resetting frequency $f_{reset}$ = 833 Hz. (Resetting pulses not shown for clarity.) 3 clicks recorded. D) Same as C but under MIR illumination of $\lambda$ = 11.3 μm and $P$ = 6.51 nW. Many clicks recorded. E) Count rate $\Gamma$ as a function of MIR illumination power and bias voltage, measured at base temperature $T$ = 3.7 K. F) Spatial map of count rate $\Gamma$ measured at of $\lambda$ = 675 nm, where counts were measured using the differential voltage signal from contacts 1-2. The techniques using for measuring the count rate and overdrawing the layout of the device are the same as for Fig. 3A in the main text. Reported powers are corrected for absorption in BLG (~4 %, see Supplementary Section 18).

### S12 – Contact dependence of negative differential conductance

The spatially resolved measurements of Fig. 3A suggest the negative differential conductance (NDC) is localised to a specific region of the device. To further test this conclusion, we performed high-field transport measurements between different pairs of contacts. Fig. S11A plots an optical image of the device with the contacts labelled. Figs. S11B-E plot the IV characteristics between different contact pairs and the insets plot the corresponding differential conductivity. Notably, we find the NDC can only be observed if the current path crosses the photoactive region mapped in Fig. 3A. These measurements further confirm that the NDC is localised to a small region of the device.

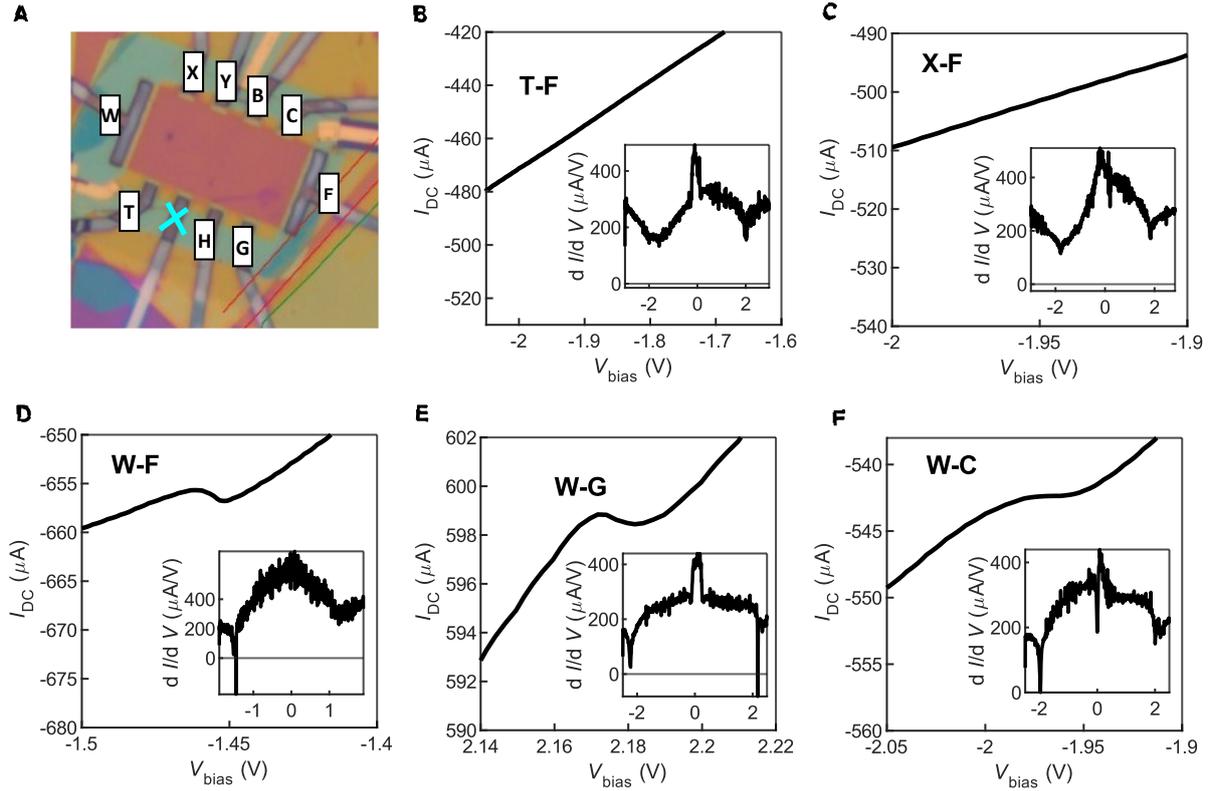

**Fig. S11 High bias IV curves in different regions of DBLG$_A$. A)** Optical image of the sample. **B-F)** insets: numerical derivative of the complete *IV* curve plotted vs $V_{bias}$. Datapoints below zero are indicative of NDC. Main panels: raw *IV* curves data in the region where d*I*/d*V* is lowest.

**Supplementary Section 13 – Two region model for NDC**

Our spatially resolved photon counting (Fig. 3, Fig. S10) and transport measurements (Fig. S11) tells us that the negative differential conductance (NDC) occurs in small regions of our devices. Hence, the overall IV characteristics of our devices is built from different regions with distinct IV characteristics, connected either in series or in parallel. This could explain why the high field transport is mostly dominated by the out-of-equilibrium criticalities[21], which were found to be commonplace in graphene-based moiré systems. To explore this interplay, we devise a two-region model which, by solving a system of differential equations, calculates the resulting IV characteristics of two non-linear elements placed in series. We approximate both regions' IV characteristics with a 3rd order polynomial $I_{1,2}(E)$ (here $E$ is the driving electric field along the plane), where $I_1(E)$ has a positive derivative and represents a region where Zener tunnelling appears before the single-band NDC would appear (Fig. S12 green curve), while $I_2(E)$ is tuned to approximate the single-band negative differential conductivity found from the kinetic equation (Fig. S12 orange curve, see Supplementary Section 7 for details on the kinetic model). For a series connection of two regions, we solve the equations: $I_1(E_1) = I_2(E_2)$ and find the total bias voltage as $V = l_1 E_1 + l_2 E_2$ where $l_{1,2}$ are the lengths of the corresponding regions and the fraction of NDC region is defined as $l_2/(l_1 + l_2)$.

Combining the two regions, we find the resulting IV curve (Fig. S12 black dashed line), which resembles our measurements in DBLG$_A$ very closely (*cf.* Fig. 1D). Specifically, the IV curve exhibits predominantly a positive differential resistance but hosts a narrow voltage range where the NDC occurs. The mixing of two regions essentially narrows the voltage range where NDC is observed. As the NDC region gets smaller (black curve), the voltage range narrows further and the critical current shifts closer to that of the non-NDC region. This may explain why our observation of NDC occurs for similar current densities to the Schwinger criticalities, because only <20 % of our device exhibits the NDC behaviour (*cf.* Fig. 3A of main text) whilst the rest exhibits positive differential resistance. These observations exemplify how the intrinsic IV curve of the negative velocity region (Fig. 4C) may be masked by the regions exhibiting Schwinger transitions. We note that changes in the critical field of the NDC regime will still influence the IV of the whole system, including changes to the critical field. Hence, our analysis of the bandwidth dependence on the critical field is still justified in a two-region model.

We also comment on the peculiar temperature dependence observed in Fig. 5, where critical voltage shifts to lower values as temperature is increased (Fig. 5A), and can be explained using the two-region model. An increase in temperature leads to increased scattering, making it less probable for the carriers to reach negative-velocity regions of the Brillouin zone. In a pure NDC region, this would mean *increasing* the necessary field, meaning a larger critical voltage. However, scattering will also make NDC less pronounced, which, in the two-region model, would render the dashed curve to become closer to the black curve, resulting in a *decreased* critical voltage.

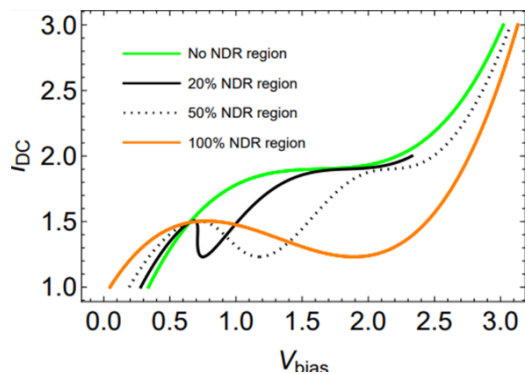

**Fig. S12 Two-region model.** Combining an NDC region (orange curve) with a non-NDC region (green) in series results in an IV curve showing a narrower NDC feature with the similar critical voltage (dashed curve). As the NDC region is shrunk relative to the non-NDC region, this trend continues (black curve) giving a curve that closely matches the non-NDC characteristics, except in the vicinity of the critical voltage where the NDC feature is sharp and narrow. B) Shifting the critical voltage of the NDC region (orange curve) results in a corresponding shift of the overall IV characteristics for combined regions (black curve).

**Supplementary Section 14 – Polarization dependence of count-rate**

Figure. 3A shows that the photoactivity is localised to a small region within the device. Our scanning photocurrent measurements (Supplementary Section 14) map this area to the bulk of the sample, but is located close to the current injectors. To rule out contact effects, we performed polarization-dependent measurements of the count-rate allowing us to disentangle the mechanism underlying the single-photon switching. Fig. S13 plots the count-rate as a function of polarization-angle rotated using a mid-infrared half wave plate. It shows no dependence on the polarization direction ruling out lightning rod effects originating from contacts and further corroborating the response originates from the bulk.

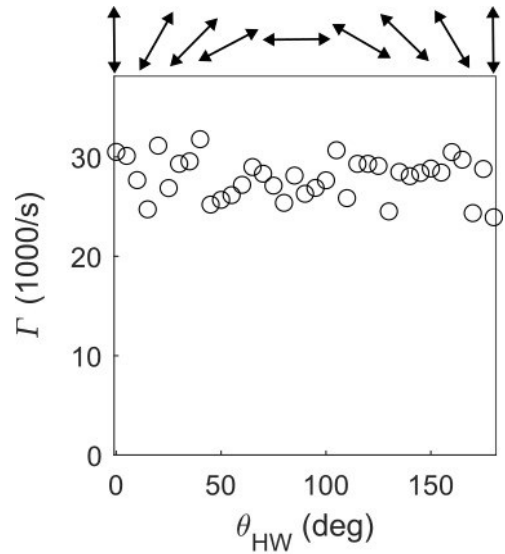

**Figure. S13 Polarization dependence on the count rate**. Count rate is plotted as a function of the half-wave plate angles (polarization direction indicated above the graph). For these measurements light is shone without the objective with absorbed powers $\sim$ 7.2 nW, and a reset frequency of 38 KHz. The system is current biased close to the reset frequency limited saturation point.

**Supplementary Section 15 – Circuit-limited reset frequency**

Operating our devices with higher resetting frequency lowers the dead time, enabling faster operation and higher detection efficiencies. As in any electronic circuit, there are however limitations restricting the maximum operating frequency achievable in our experiment. These limitations may be intrinsic to the device operation, or extrinsic, i.e., related to the set-up. While testing the frequency dependence of the count rate (Fig. S14A), we noticed a cut-off at around 45 kHz (green curve), in which the system has no time to roll-off to the 'active state' voltage level between two subsequent reset pulses. We studied the extrinsic limitations of our setup by placing a dummy resistor of 2kOhm resistance (comparable to the sample resistance of 2.15 kOhm). By analyzing the temporal profile of pulse propagation through the circuit, we found that the functioning upper limit of approximately 38 kHz comes from the extrinsic RC time constant of our setup (around 7 µs), rather than from the device itself. The RC constant is extracted from looking at the rate of fall off of the voltage after the resetting pulse (Fig. S14B), according to the definition of the time constant for a capacitor. In general, electronics systems need around 4-time constants to recover to the base level after excitation by a pulse and indeed we observe that 4/ (7 µs) = 35.7 kHz is around the maximum reset rate we can use. According to simulations done with QUCS software, the time constant of the circuit used in our experiments should be around 1 µs, coming mostly from the parasitic capacitance of the cables and leads which are not impedance matched to the device's resistance. To test this simulation, we performed a control experiment comparing the signal propagation time through our sample and the same test resistor using a circuit that supposedly replicates that of our cryostat (Figure S14C). In this configuration, we measured an RC time of 1.5 µs for both the device and test resistor, close to the simulated values. One possible source of the longer time constant in the cryostat may originate from additional parasitic capacitance which is not considered in the simulation, like DC current source, imperfect connections, soldering issues, etc.

In general, for NDC devices the limiting speed is given by the capacitance of the device and the amplitude of the flowing current. The calculated capacitances of the sample (0.4 pF for the Hall bar and 82 pF for the un-optimized gold leads layout) are small compared to the wiring capacitance (750 pF as estimated from Fig. S14C, *i.e. outside* of the cryostat) meaning that the maximum detector's reset rate can be orders of magnitude higher than 38 kHz, if it is operated using a dedicated electronic circuit designed for this application.

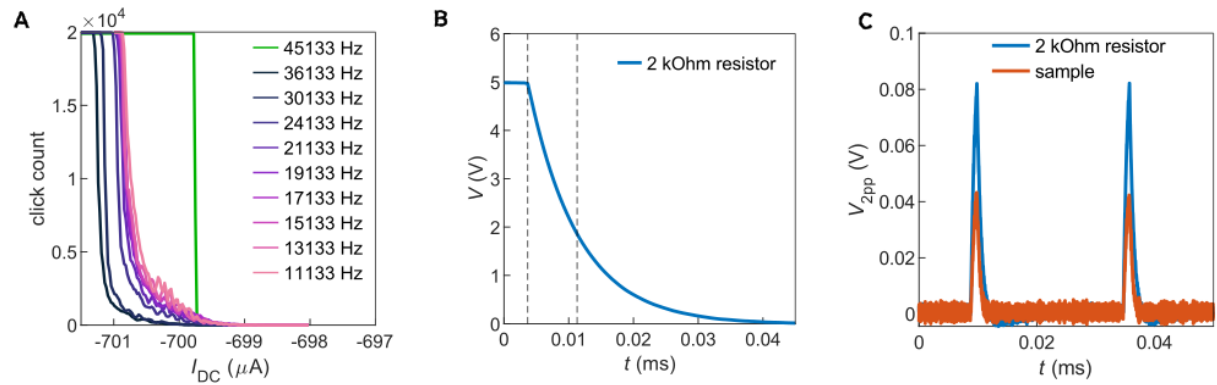

**Figure. S14 Speed. A)** The effect of changing the reset frequency on the count rate's dependence on bias (normalized by integration time for fair comparison). The curve gets steeper with increasing $f_{reset}$ which is beneficial for photodetection, however if $f_{reset}$ is too high the system cannot properly reset (green curve shows an abrupt jump from 0 counts to saturation). **B)** The fall of a 5 V voltage pulse over time for a 2 kOhm resistor placed instead of the sample in our cryostat (at room temperature). The dashed lines mark the 7 μs time interval over which the voltage drops to 1/e of the maximum value. **C)** Comparison of the time traces measured outside of the cryostat, plotting a voltage pulse propagation through our sample (orange) and a test resistor (blue), showing a fall time of 1.5 μs.

**Supplementary Section 16 – Exponential dependence of count rate**

As mentioned in the main text, the exponential dependence of count rate on bias is a hallmark of probabilistic detection in superconducting nanowire single-photon detectors (SNSPDs). This means that the absorption of a photon has a smaller-than-unity chance of causing a detector click, since these events need to be accompanied by a temperature-induced fluctuation and are themselves subject to Fano fluctuations[22]. In this section, we study the exponential dependence of count rate ($\Gamma$) to learn more about the switching mechanism.

Fig. S15A plots $\Gamma_{dark}$ as a function of current bias ($I_{DC}$) for three different temperatures ($T$), exhibiting a sharp rise of dark count rate $\Gamma_{dark}$ with $I_{DC}$ for all $T$. In Fig. 15B, we plot the natural logarithm of these three curves as a function of $\Delta I_{bias} = I_{DC} - I_{sat}$, where $I_{sat}$ corresponds to the onset of saturation region (*i.e.* $\Gamma_{dark} = f_{reset}$) at a given temperature. Two observations are striking.

First, all the three curves overlap very closely, meaning that increasing the temperature only causes a shift of the $\Gamma_{dark}(I_{DC})$ and, consequently, the critical current, but does not affect its shape nor slope. Second, $\Gamma_{dark}$ shows an exponential dependence on bias with two distinct regimes, denoted as I and II and shaded by green boxes. Far away from $I_{sat}$, $\Gamma_{dark}$ grows exponentially with bias current until a cross-over at around $\Gamma_{dark} = 0.05\ f_{reset}$ where the exponent value suddenly changes. The dependence stays exponential until it breaks down at $\Gamma_{dark} = 0.75\ f_{reset}$ as we approach detector saturation.

We note that such a behaviour can be related to the bistable nature of our detector. Bistable systems are characterized by an energy barrier $U$ separating two local minima in the potential energy of a system. *E.g.* Fig. 15C depicts a so-called washboard potential in a biased Josephson junction, adapted from ref. [23]. The system (represented by the dot) can escape the local minimum either via thermal activation (TA) or through macroscopic quantum tunnelling (MQT)[24], and increasing the bias current allows one to control the relative contribution of these mechanisms to the total count rate.

To understand further the behaviour of the potential barrier *U*, we analyse the temperature dependence of the dark count rate. Fig. S15D plots the natural logarithm of $\Gamma_{dark}$ as a function of the inverse temperature $T^{-1}$ (Arrhenius plot) for a few different bias values. Interestingly, the dependence of $\Gamma_{dark}$ on $T^{-1}$ is also exponential and, similarly to bias dependence (Fig. S15B), there are two regimes distinguished by a different exponent. Considering the exponential temperature dependence, however, has the advantage that we can assume a well-known thermal activation model and extract the barriers' heights from the linear fits. According to thermal activation model:

$$\Gamma_{\text{dark}} = A \cdot \exp\left(-U(I_{\text{DC}})/(k_{\text{B}}T)\right),$$

hence

$$\ln\left(\Gamma_{\text{dark}}\right) = \ln\left(A\right) - U(I_{\text{DC}})/(k_{\text{B}}T).$$

Comparing this to the linear fit in the Arrhenius plot

$$\ln(\Gamma) = a/(kT) + b,$$

it is clear that

$$a = U(I_{\text{DC}}), b = \ln(A).$$

Motivated by the observation that temperature does not affect the shape or slope of the curves in Fig. 15B, we infer that *U* is temperature independent. This validates the use of the thermal activation model to extract the value of *U* at every $I_{DC}$ from the Arrhenius plot. The extracted values of *U* for the two different regimes are shown in Fig. S15E. Apparently, the height of the barrier changes significantly with the bias in the two regimes, reaching $U_1 \approx 18$ meV in regime I and as high as $U_2 \approx 60$ meV in regime II.

Based on the above evidence, we conclude that there are two mechanisms responsible for triggering dark counts in our detector, and which one is dominating depends on the bias current relative to the critical current $I_C$.

Finally, we turn our attention to the power dependence of the exponential trend. Fig. S15F is essentially the same as Fig. S15B but for varying illumination powers instead of a varying temperature. In a striking contrast to the temperature dependence, the traces do not overlap, meaning that illumination is influencing the fitting parameters *a* and *b*. Interestingly, even though the shift of the saturation regime was greater in the temperature dependence in the tested ranges (*cf.* Fig. 2E and Fig. 2F of the main text), the change of the slope is much more significant in the power dependence. It is evident that MIR illumination is affecting $\Gamma$ qualitatively differently from temperature even before the detector reaches saturation, providing strong evidence that the increase of switching rate is not related to photon-induced heating.

The microscopic mechanism responsible for the switching remains unknown and requires further studies. that are beyond the scope of this work. For the sake of context, we would like to note that up to this day, two decades from the first demonstration of a working SNSPD, the exact microscopic mechanism leading to breaking of superconductivity upon photon absorption in these devices is still being debated[22]. This has, however, not impeded the rapid advancement of this technology and we expect that the same should hold true for MSPDs.

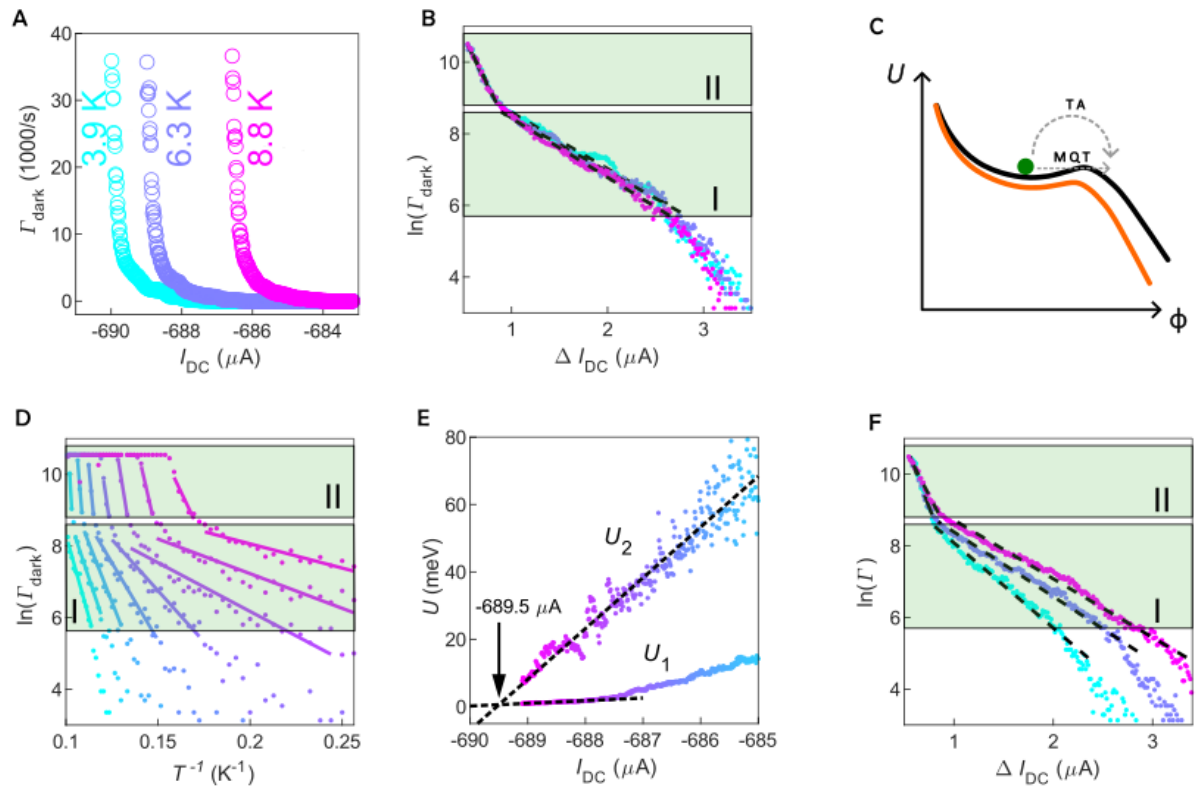

**Fig. S15 Exponential scaling.** A) $\Gamma_{dark}$ versus $I_{DC}$ for three different temperatures (horizontal cross-sections through Fig. 2E). B) the same data as in (a) but on a semi-logarithmic scale and shifted so that the leftmost points overlap. Two distinct linear regimes are evident, marked by green boxes. Increasing the temperature does not change the shape or slope of the curves. C) The dependence of potential energy $U$ on the phase $\phi$ for two different values of bias current $I_b$ in a superconducting Josephson junction photodetector. The system can jump out of the local minimum either by a thermal activation process (TA) or macroscopic quantum tunneling process (MQT) D) Arrhenius plot of $\Gamma$, showing $\ln(\Gamma)$ versus inverse temperature for a few different values of $I_{DC}$. Also here, two distinct linear regimes are evident (marked by green boxes). The bias values are color-coded according to the bias values in panel E. E) Barrier heights extracted from thermal activation model in two regimes, plotted together against $I_{DC}$. F) Natural logarithm of $\Gamma$ versus $\Delta I_{DC}$ for three different powers (0 nW, 7.2 nW and 14.4 nW, from teal to pink). Contrary to the temperature (panel B), laser light induces a change of slope, showing that the increase of the count rate due to illumination does not originate from laser-induced heating of the crystal lattice.

**Supplementary Section 17 – Gate dependence of photoactive region**

In Fig. 3A of the main text, we showed spatially resolved measurements of the single-photon counting where the photoactive region lies in a narrow strip of the device. This optically active region also tells us that the NDC is localised to a specific region of the device. In Supplementary Section 3, we discussed the prominence of self-gating effects in our devices. In extreme cases, self gating can lead to pinch-off effects similar to those reported previously in monolayer graphene devices[25]. If the the photoactive region was created by a pinch-off, it should move spatially depending on the gate-voltage and/or bias voltage that is applied. Fig. S16B,C plots spatial dependent measurements of the count rate measured for positive (**B**) and negative (**C**) current densities at which the NDC occurs. In both maps the photoactive region remains fixed between $X = 130 - 140$ μm. Fig. S16D-F plots measurements for different doping levels where NDC could be observed. Notably, regardless of the doping level, the photoactive region remains fixed. Fig. S16 hence provides evidence that local nature of the NDC is not induced by self-gating/pinch-off effects but rather related to material structure.

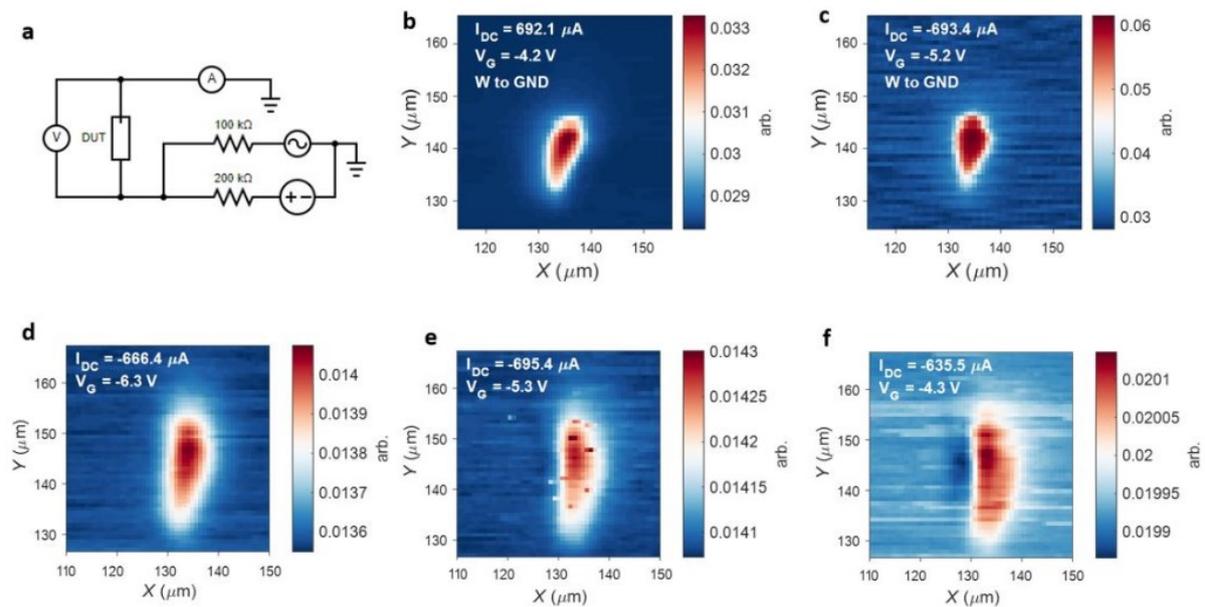

**Fig. S16 A)** Circuit diagram for the measurement scheme used for spatially resolved single-photon measurements. **B – C)** $\Gamma_{arb}$ plotted as a function of position X,Y for a fixed gate voltage ($V_G$) and positive (b) and negative (c) currents. **D – F)** Measurements performed for different $V_G$ and $I_{DC}$ close to the critical current $I_c$.

**Supplementary Section 18 – Bistabilities observed in magic-angle twisted bilayer graphene**

We have also observed signatures of bistability in a large-area magic-angle twisted bilayer graphene device. Figure S17A shows an optical image of the studied device. Fig. S17B plots the two-probe resistance measured between the wide current injecting contacts as a function of the gate voltage ($V_{gate}$). We observe that $R_{2pp}$ exhibits a peak close to the charge neutrality point (CNP) ($V_{gate}$ = 0) and two satellite peaks indicating full filling of the moiré Brillouin zone. From the position of these peaks and Hall effect measurements we extract a twist-angle of 1.18°. The overall high resistance and oscillatory structures provide additional evidence for flat-electronic bands and correlated insulating states[26]. For most of the carrier doping we observed linear IV characteristics resembling previous reports. However, fixing the doping in the remote bands beyond full filling, (red arrow in Fig. S17B) we observed sharp jumps in the IV characteristics (Fig. S17C). The jumps signify bi-stabilities we believe originates from NDC. However, in our magic-angle samples the contact resistance (approximately 20 kOhms) is significantly larger than our BLG devices (approximately 2 kOhm), , which prevents us from observing NDC as a large resistance in series collapses the NDC region to a sharp jump (see supplementary section 2). Further work is required to understand the key design principles for observing NDC and bistabilities in twisted bilayer graphene. Nonetheless, Fig. S17 demonstrates the physics underlying is commonplace to different moiré superlattice systems.

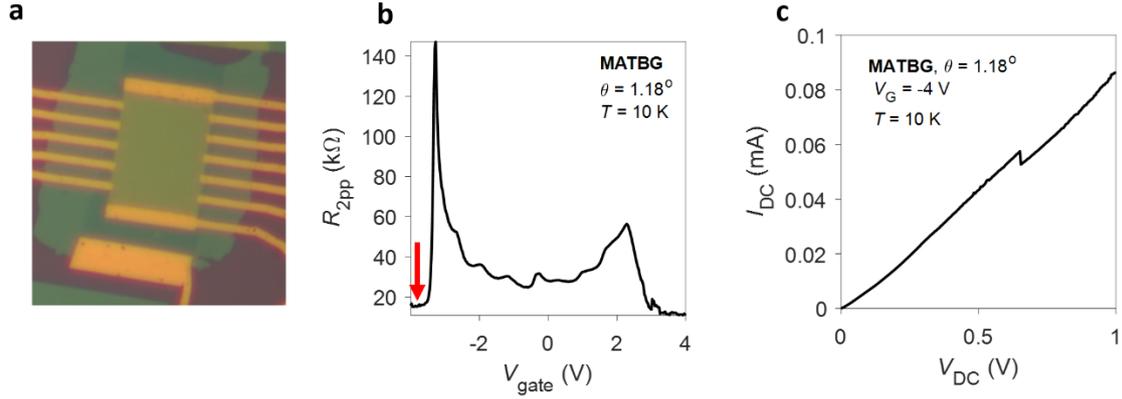

**Fig. S17 A)** optical image of the studied device. **B)** two-probe resistance ($R_{2\text{-probe}}$) measured between wide current injector contacts, plotted as a function of the gate voltage $V_{\text{gate}}$. Temperature ($T$) is 10 K and the twist-angle ($\theta$) extracted from Hall effect measurements is 1.18°. **C)** current ($I_{DC}$) measured as a function of voltage ($V_{DC}$) sourced through the injector contacts. The doping level is depicted by the red arrow in (B).

**Supplementary Section 19 – RCWA calculations of absorption spectra**

In our two-dimensional material heterostructure, only a small fraction of the incident photons is absorbed due to its inherent few-layer thickness.[27] While near-unity absorption can be achieved using appropriate device engineering[28], in our proof-of-principle device the absorptance is much smaller than unity. Therefore, to correctly quantify the device performance and compare with other single-photon detectors built from bulk materials, we report the power values normalized by the fraction of absorbed photons. For this, Rigorous Coupled Wave Analysis (RCWA) was used, following the approach applied by Agarwal *et. al* for similar structures[29]. In summary, this method determines the absorption coefficient by solving the Maxwell's equations for incoming plane waves. Our structure of interest was a BLG layer embedded in a heterostructure, which includes hexagonal-boron nitride layers and graphite gates. Inside each layer, the in-plane fields are expanded via Fourier transforms, and boundary conditions are used to connect the solutions. This provides the scattering parameters and electric field distribution, allowing the full optical response to be computed. The optical conductivity for each material is derived based on known properties[29], and the absorption spectrum is then reconstructed from the scattering parameters.

Here, we consider a simple BLG as an approximation of our BLG/hBN superlattice system, since the exact calculation of the absorption spectrum of BLG/hBN superlattice is non-trivial and out of the scope of our work. We consider four bands of BLG because the remote bands may play a role when visible wavelengths are used. Fig. S18A displays the simulated structure corresponding to DBLG$_A$ which consists of many encapsulating layers of hexagonal-boron-nitride, and two layers of graphite gates.

Fig. S18B plots the calculated absorption in the BLG layer as a function of the wavelength in our range of interest. At $\lambda$ = 675 nm the absorption is approximately $\alpha$ = 2%, while at $\lambda$ = 11.3 µm it is around $\alpha$ = 3%. These values were the ones we used for calculation of the illumination power densities $P$ according to the formula $P = \alpha P_{\text{tot}}$, where $P_{\text{tot}}$ is the average power density incident on the sample area. $P_{\text{tot}}$ was obtained from measuring the total power in the lasers' beams, multiplying it by the measured transmission of the objective and cryostat windows and dividing by the ratio of the focal spot area (deduced from the photovoltage spatial maps) to the upper limit of the active area size (*cf.* Fig. 3A of the main text). The measurement of the windows' transmission was performed with the windows placed after the objective, so the light impinging on the window was covering all the incidence angles allowed by the objective's NA.

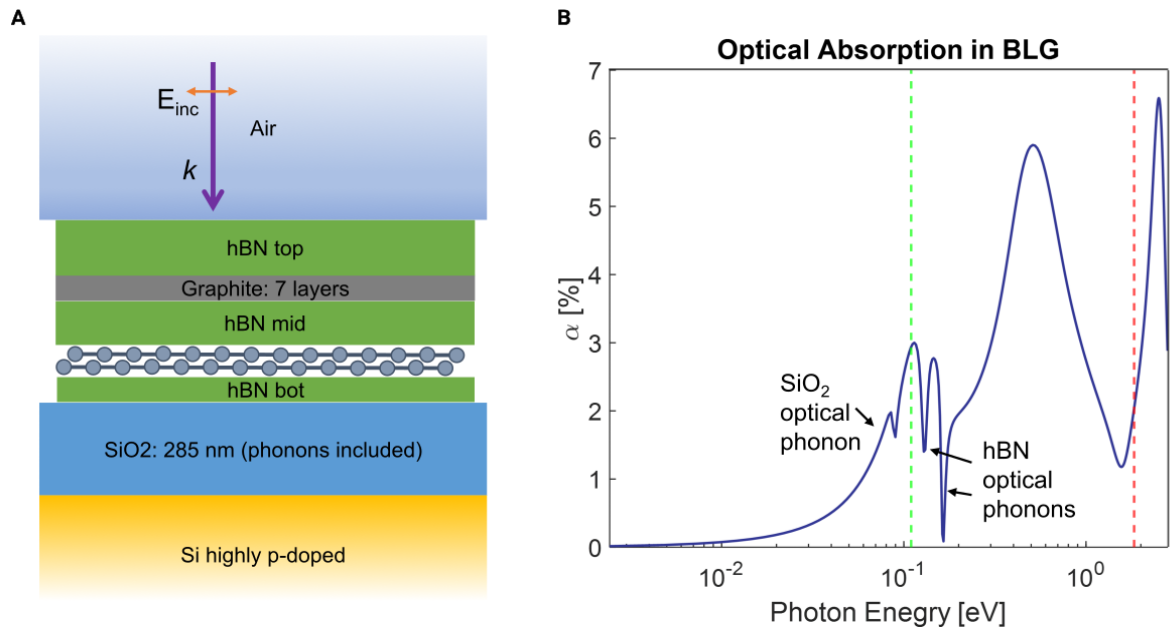

**Fig. S18 A**) Illustration of the simulated 2D materials stack, representative of the DBLG$_A$. Alignment between hBN and BLG is not included in the model. **B**) The absorption (α) in the bilayer graphene embedded in the stack shown in panel (A), as a function of photon energy. The dashed red and green lines correspond to the absorption values at λ = 675 nm and λ = 11.3 μm, respectively.